\documentclass{nature}
\usepackage[utf8]{inputenc}
\usepackage[T1]{fontenc}
\usepackage{subfig}

\usepackage{adjustbox}
\usepackage{amsmath}
\usepackage{amssymb}
\usepackage{multirow}
\usepackage{lscape}
\usepackage{booktabs}
\usepackage{graphicx}% Include figure files
\usepackage{tabularx}
\usepackage{array}
\usepackage{makecell}
\usepackage{multirow}
\usepackage{color}
\usepackage{xcolor}
\usepackage[printwatermark]{xwatermark}
\usepackage{xspace}
\usepackage{caption}

\title{Academic Support Network Reflects Doctoral Experience and Productivity}
\author{Ozgur Can Seckin,$^{1}$ Onur Varol$^{1,2,\ast}$}

\begin{document}

\maketitle
	
	\begin{affiliations}
		\item Faculty of Engineering and Natural Sciences, Sabanci University, Istanbul, 34956, Turkey
		\item Center of Excellence in Data Analytics, Sabanci University, Istanbul, 34956, Turkey
		
		\normalsize{$^\ast$To whom correspondence should be addressed; E-mail:  onur.varol@sabanciuniv.edu.}
	\end{affiliations}

    \begin{abstract}
    Current practices of quantifying performance by productivity leads serious concerns for psychological well-being of doctoral students and influence of research environment is often neglected in research evaluations. Acknowledgements in dissertations reflect the student experience and provide an opportunity to thank the people who support them.
    We conduct textual analysis of acknowledgments to build the ``academic support network,'' uncovering five distinct communities: Academic, Administration, Family, Friends \& Colleagues, and Spiritual; each of which is acknowledged differently by genders and disciplines.
    Female students mention fewer people from each community except for their families and total number of people mentioned in acknowledgements allows disciplines to be categorized as either individual science or team science.
    We also show that number of people mentioned from academic community is positively correlated with productivity and institutional rankings are found to be correlated with productivity and size of academic support networks but show no effect on students' sentiment on acknowledgements.
    Our results indicate the importance of academic support networks by explaining how they differ and how they influence productivity.
    
    \end{abstract}

\thispagestyle{empty}

\section*{Introduction}

In recent years, well-being and mental health concerns for PhD students have been increasing.  
According to a recent survey conducted in 2019 by Nature on 6,300 PhD students, 36\% responded that they sought help for anxiety or depression caused by their studies.\cite{woolston2019phds} 
Another devastating fact is that doctoral students are 2.43 times more likely to have a common psychiatric disorder than the rest of the highly educated population.\cite{levecque2017work}
It is therefore important to look through the journey of doctoral students not only through the lens of academic ``success measures'' such as publication numbers, citation counts, fellowships received \textit{etc.} but also at their overall well-being and the quality of the environment that supports them in fulfilling their potential.

Although obtaining a doctoral degree is often viewed as an isolated process, it is a collaborative endeavour in which family, friends, colleagues, advisors, faculty, and administrative staff are directly or indirectly involved and can influence the well-being of the students. 
At the end of the journey, students can show their gratitude by mentioning these names in their work through ``acknowledgements'' section of their dissertations. 
Acknowledgements, even though existed before, could not be found explicitly in the academic work before 1940s, and did not become a common subsection until 1960s.\cite{bazerman1988shaping}  
Hyland named acknowledgements as ``Cindirella'' genre because of its suffering from an undeserved neglect.\cite{hyland2003dissertation}
Through time, these sections got longer and their use have become more prevalent,\cite{cronin1992patterns} making them more ``insightful'' in terms of understanding how and with whom doctoral students complete their journeys.
Since there is almost no guideline or style guide to receive help when writing this section,\cite{hyland2004graduates} students have more freedom, compared to the other parts of their dissertations. 
Acknowledgements also serve purposes other than expressing gratitude, such as exhibiting associations with respected academics to display a special connection to which the author has been admitted.\cite{scrivener2009exploratory} 
Thus, introducing their strategic decision in their professional identities by illustrating the author in a positive aspect and governing their connections with the disciplinary community.\cite{ben1987acknowledgements}

Acknowledgements contain such profound details of their authors' academic journey; however, research efforts to study how they vary concerning disciplinary and demographic differences have remained limited. Mantai and Dowling examined the type of social support that are provided for PhD students using 79 acknowledgements gathered from Australian universities.\cite{mantai2015supporting} 
Hyland examined 240 acknowledgements of MA and PhD dissertations to characterize their narrative structure.\cite{hyland2004graduates}

Using acknowledgement sections to delve into hidden networks outlined by the gratitude and appreciation expressed by students helps drawing conclusions that cannot be obtained from measures of academic success alone.
For this task, we examined 26,236 PhD dissertations, obtained from \textit{ProQuest Open Access Dissertations \& Theses} database (PQDT-Open hereafter), 99\% of which are from the United States in the last 20 years.
We aimed to shed light on the doctoral process by examining who is acknowledged, and how they are recognised from the perspective of students, using the tools of network science and natural language processing that enable research on large-scale data.
We revealed gender based and disciplinary differences when acknowledging support providers in terms of number of people mentioned and sentiment scores.
We also investigated the factors derived from academic support networks influencing productivity levels.
Lastly, we point out to linguistic differences between those who are located in the extreme cases of productivity and and of sentiment.

\section*{Results}

\subsection{Characterization of support network}
The acknowledgements section of dissertations contains statements about individuals or institutions who have provided emotional, economic, and administrative support to students on their journey towards attaining their degree. To systematically identify acknowledged individuals and institutions, we used a data-driven approach supported by manual inspection to identify distinct types of support providing entities in the acknowledgements.

\begin{figure}[htp]
    \centering
    \includegraphics[width=0.8\linewidth]{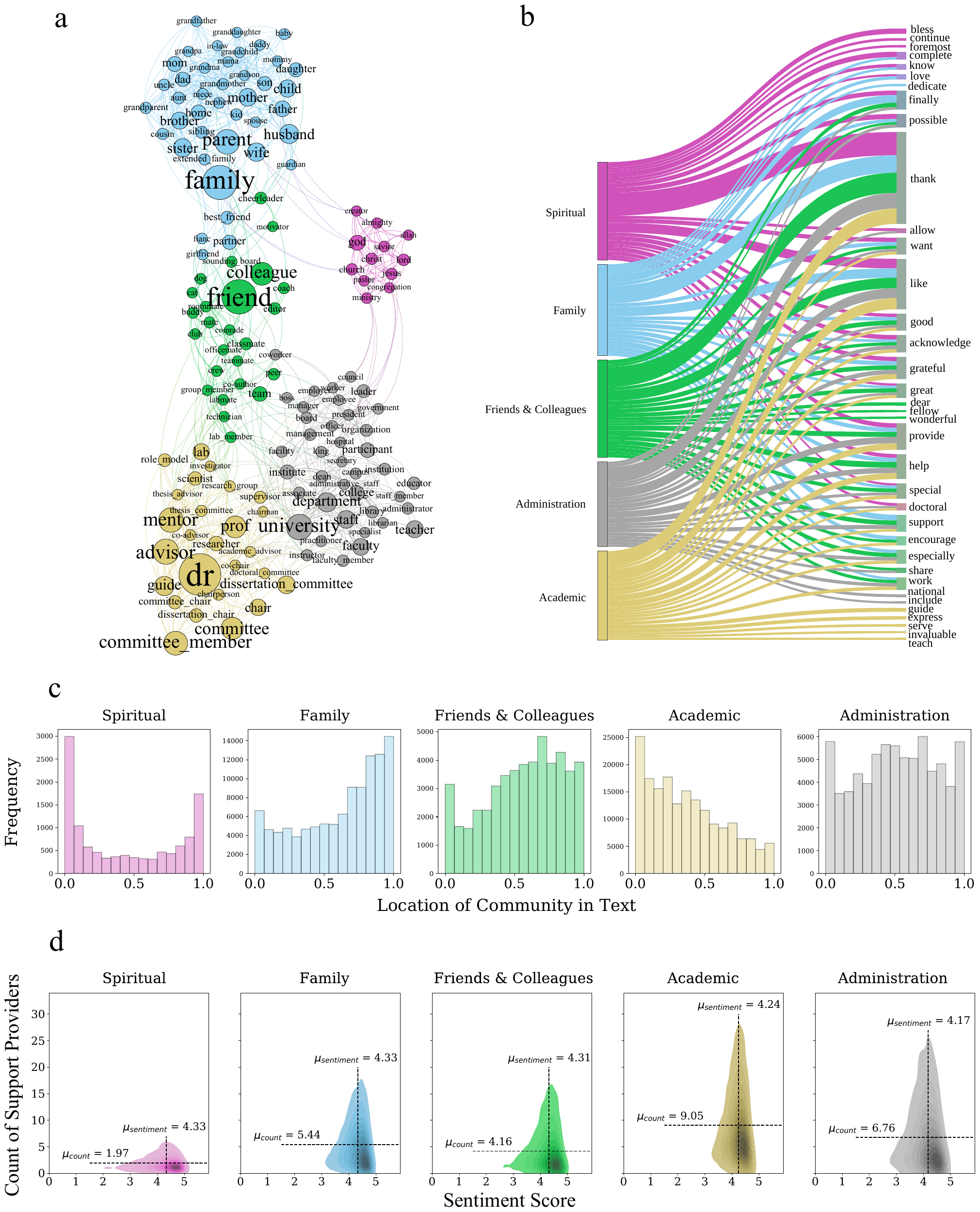}
    \caption{\textbf{Analyzing support providers in acknowledgements.} Different support providing entities identified in dissertation documents represented as nodes and their contextual similarities learned from document embeddings used as edge weights (a). Community detection revealed 5 distinct groups: Family, Friends \& Colleagues, Academic, Administration and Spiritual. These groups are acknowledged using specific words and bi-partite relation points group specific properties (b). Location of mention in the acknowledgement text indicates norms among scholars to highlight distinct groups (c). Support providers are also differ in terms of their occurrence in acknowledgements and the corresponding sentiment they are referred (d).}
    \label{fig:support-network}
\end{figure}

To build the academic support network, we extracted different individual roles and institution types as nodes from each text and computed contextual similarities learned from the text as edge weights (Fig.\ref{fig:support-network}(a)).
Our entity extraction approach identified 144 support providers that were mentioned in at least 50 dissertations.
We used a deep learning approach, called Doc2Vec,\cite{le2014distributed} to learn embeddings for each support provider within the context they were used in the dissertation corpus. Using these embeddings, we calculated similarities as edge weights between embeddings learned for each node. We used disparity filtering and retained only the statistically significant edges (see SI:Sec-\ref{secSI:creating_a_network}), therefore giving us the network that captures significant relations between these entities.
Later we employed Girvan-Newman\cite{girvan2002community} algorithm for community detection to identify groups of support providers.

The network representation of all of the support providers is given in Fig.\ref{fig:support-network}(a).
Community detection analysis identified 5 distinct communities in this network and each of them is illustrated by a different color: Spiritual (purple), Academic (yellow), Administration (gray), Family (blue) and Friends \& Colleagues (green). These communities are consistent with those identified with other clustering approaches like hierarchical clustering as well (SI:Sec-\ref{secSI:creating_a_network}).
Node sizes were determined by the occurrence of support providers and the edges were weighted with cosine similarity of embeddings between node pairs.

Connectivity among these communities reveals separation between social and professional networks . Friends \& Colleagues are located among Family, Academic and Administration communities. Some dissertations also refer to spiritual entities and community consisting of these entities is loosely connected with the rest of the network and has few links with the family community.
Factors influencing community relations can be explained by comparing the words that are used to acknowledge these communities.
By analyzing bipartite connections of support providers and prominent words, as seen in Fig.\ref{fig:support-network}(b), we present the most frequent 20 words used for support providers in these communities. While four of the most widely used words for acknowledging Spiritual support providers are not linked with the other communities; words like thank, acknowledge, and grateful used approximately at the same rate for each group.

Hyland argued that the structure of dissertation acknowledgements has, in general, a ``thanking move'' section, in which authors start by presenting the participants, continuing next by thanking them for academic assistance (i.e. intellectual support, ideas), then for resources (i.e. technical, financial support) and lastly for moral support (i.e. friendship, patience).\cite{hyland2004graduates} In our academic support network, we observed a similar narrative structure. 
To further support rank order in acknowledgements, we checked the locations of the support providers in the text and observed that different communities can be distinguished by analyzing the locations in which they are frequently mentioned (Fig.\ref{fig:support-network}(c)).
While academic support providers are most frequently mentioned at the beginning of the acknowledgements, students tend to start talking about their families towards the end.
We also observed that Friends \& Colleagues and Administration are generally mentioned in the middle of the text. In contrast, Spiritual entities are mentioned either at the beginning or at the end. 

Although acknowledgements are expected to have an overall positive sentiment, certain entities receive more formal tone. To highlight these subtle differences, we explored the interplay between sentiment scores and how many times they are mentioned for each category separately as shown in Fig.\ref{fig:support-network}(d).
The sentiment analysis results have shown that Spiritual, Family and Friends \& Colleagues communities are being acknowledged roughly at the same level with average sentiment scores 4.33, 4.33, and 4.31 defined in $[0,5]$ range. Academic and Administration communities have lower scores on average 4.24 and 4.17, respectively.
Similarly, we analyzed the number of people mentioned from these categories. Not surprisingly, PhD students tend to acknowledge the Academic community the most (9.05 people per acknowledgement), it is followed by Administration (6.76), Family (5.44), Friends \& Colleagues (4.16), and Spiritual (1.97).
Families, friends, and spiritual figures generally do not involve in research as workforce; however, they provide emotional and financial support to make life easier for doctoral students and are a crucial part of the support network deserving an appropriate mention.

\textbf{Gender based differences}
PQDT-Open provides metadata on universities, authors, and committee but lacks details about demographics of the authors such as gender of the author. We inferred this information using the names with a widely used public API and examined the differences between genders in terms of their academic support networks (see SI:Sec-\ref{secSI:gender_inference}).
Previous work studied how female and male students acknowledge support providers both in quantitative and qualitative terms. 
Alotaibi, using Metadiscourse, studied 120 dissertation acknowledgments written by Saudi students at U.S and revealed that while all male and female students acknowledge their academic environment, there exist differences when thanking God, resources and moral support.\cite{alotaibi2018metadiscourse}
It is also shown that women in academia have less access to powerful social networks and inter-personal bounds that provide resources and create other advantages, which limits their opportunities to achieve their goals.\cite{genderinequalitycasad,HiddenPatternsCollins}

\begin{figure}[htp]
    \centering
    \includegraphics[width=16.5cm]{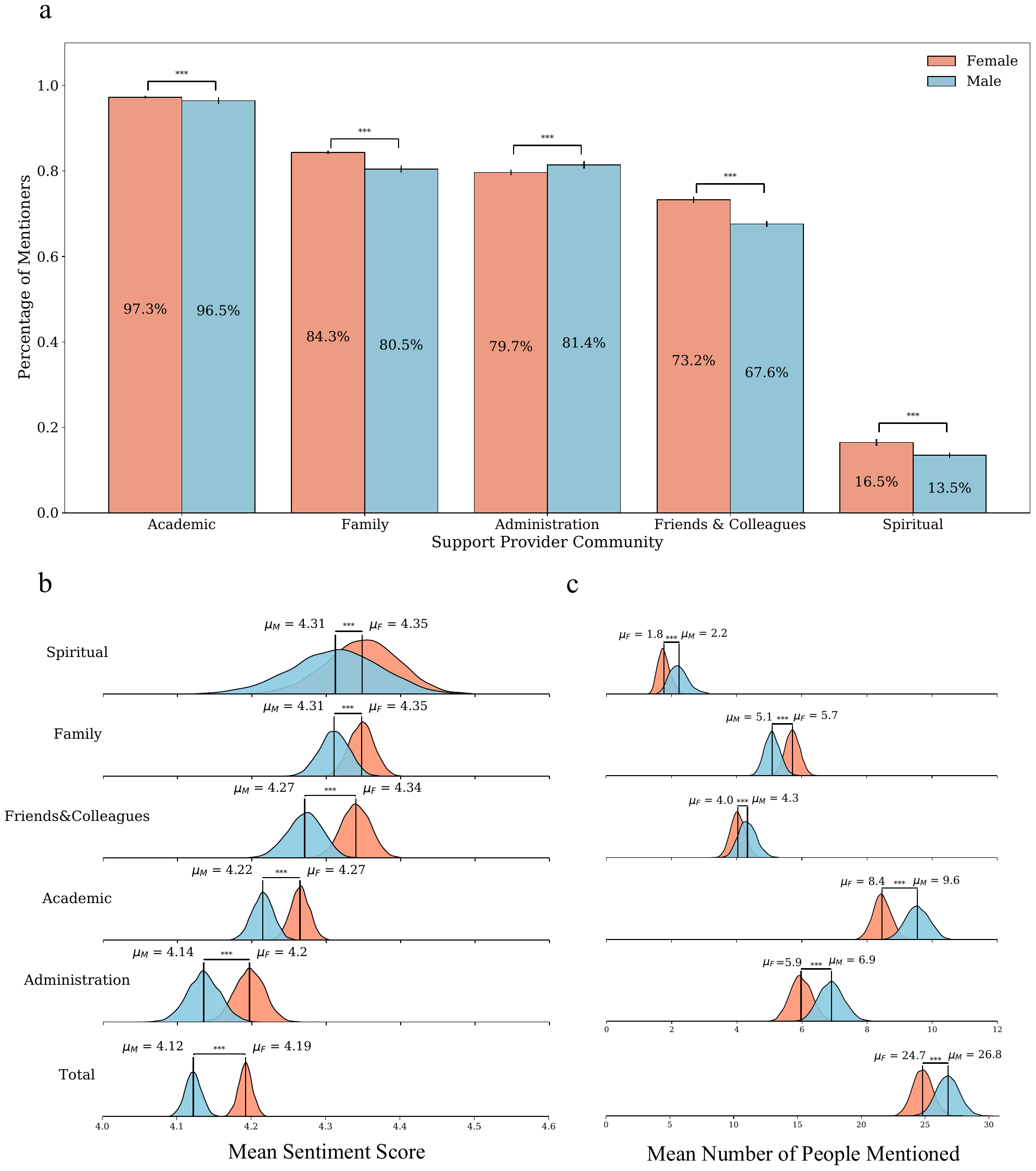}
    \caption{\textbf{Gender differences in support provider communities.} Ratio of students referring to different categories of support providers varies across genders (a). Sentiment scores differ when mentioning the support providers (b). Mean number of people mentioned alter across genders for different support provider categories (c). Individual groups were compared using the two-sided t-test. ***, $p \leq 0.001$; $**, p \leq 0.01$; $*, p \leq 0.05$}
    \label{fig:genderdifferences}
\end{figure}

Fig. \ref{fig:genderdifferences}(a) shows the ratio of students who mention the respective support provider community at least one time in acknowledgements. We observed that female students are slightly more likely to thank each community at least once except for the administration.
The largest gap is observed in Friends \& Colleagues category where the difference is 5.6\% between males and females. This is followed by a roughly 4\% difference for the family members.
These results differed when we examined the number of mentions instead of percentage of mentions; number of people acknowledged from each category is higher in male students except for the family members. In fact, the highest difference is observed in Academic, Administration, and Friends \& Colleagues groups, which may imply that women have limited access to their academic advisors, administrative staff, and peers.

Besides the occurrence rates, we analyzed the sentiment of the language used for different support providers. Females are inclined to express more positive sentiment towards the ones that help them through their journey. Meanwhile, the gender gap between Friends \& Colleagues community seem to be highest; the difference is narrower for the Spiritual characters, but still significant despite the large variance of the distribution.

Looking from an overall perspective, it is readily apparent that females tend to thank their families both qualitatively and quantitatively more compared to males. This is in line with the existing work on dissertation acknowledgements showing that while both men and women appreciate social support evenly, they highlight different aspects of it; men value companionship and collegiality, women note emotional support.\cite{mantai2015supporting}
Taken together, this may be an indicator of the level of importance of families for females and lack of professional support from the other communities during their doctoral journey.

\begin{figure}[htp]
    \centering
    \includegraphics[width=16.5cm]{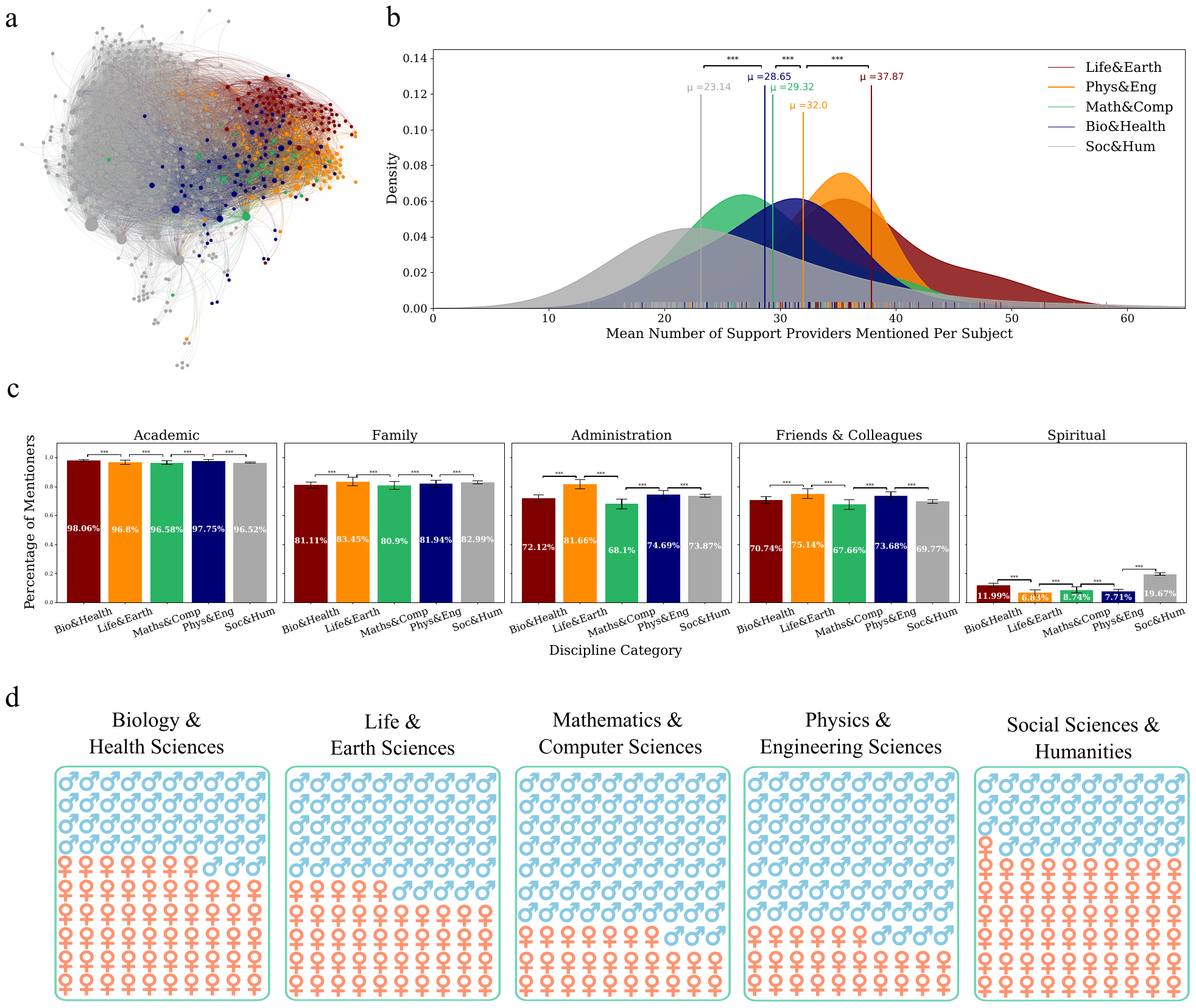}
    \caption{\textbf{Disciplinary differences.} Dissertation subjects are represented as nodes and the edges are formed by number of co-occurrences in the same document. This subject network reveals classification of field and the corresponding disciplines (a). Average number of support providers differ by disciplines ranging from individual to team sciences (b). Inclination to mention different support categories slightly diverge based on the discipline (c) and gender distributions observed to vary across disciplines (d). Individual groups were compared using the two-sided t-test. ***, $p \leq 0.001$; $**, p \leq 0.01$; $*, p \leq 0.05$}
    \label{fig:disciplinary-differences}
\end{figure}

\textbf{Disciplinary differences}
Each academic discipline has different research practices and collaborations. 
These differences are also reflected by mentorship styles and academic environment.
To analyze these disciplinary differences, we manually categorized the subjects given in the PQDT-Open metadata into five main category of academic disciplines (Biology \& Health Sciences, Life \& Earth Sciences, Mathematics \& Computer Science, Physics \& Engineering Sciences, Social Sciences \& Humanities) (see SI:Sec- \ref{secSI:discipline_categories}). Fig.\ref{fig:disciplinary-differences}(a) shows the subject co-occurrence network and categories labeled as different disciplines.

We argued that while dissertations on certain subjects may be considered as individual work and require less academic collaboration and administrative support, other subjects might require cooperation, teamwork, and access to resources and field work.
We calculated the number of support providers mentioned for each subject and presented in Fig.\ref{fig:disciplinary-differences}(b). 
While Social Sciences \& Humanities students mention the least number of people with 23.14 on average, this number is the highest with 37.87 for Life \& Earth Sciences students.
Number of support providers mentioned for each discipline aligns with academic norms of individual and team science as shown in the literature.\cite{fortunato2018science} Here, we measured not only size of academic groups, but also other support provider categories.

Moreover, it is also reasonable to presume that different disciplines might have different preferences in terms of acknowledging the support provider categories. While the results show that there is a small gap between occurrence rates, this ratio varies the most for Administration and Spiritual communities (see Fig.\ref{fig:disciplinary-differences}(c)). Mathematics \& Computer Science students seem to mention their families, friends, colleagues, and the administration less than other disciplines. Additionally, almost one fifth of Social Sciences \& Humanities students acknowledge spiritual characters, which may be explained by dissertation studies in religion and relevant fields (e.g. ``Biblical Studies'', ``Islamic Studies'') covered under this discipline.
We also investigated the gender ratios in these disciplines and, consistent with the past work,\cite{Huang4609,way2016gender} we observed that female students are underrepresented in STEM fields. Female ratio is the lowest for Physics \& Engineering Sciences with only 26\%, which is followed by 27\% in Mathematics \& Computer Sciences. However, the majority of students in Social Sciences \& Humanities are women, with a rate of 71\%. These outcomes are also supported by previous work on intersectional inequalities, where it is shown that there is homophily between identities and subject of research.\cite{kozlowski2022intersectional}

\textbf{Social determinants of the academic productivity}
Research on academic performance and success focuses on metrics that are easy to quantify, accessible for research, and standardized across disciplines.\cite{fortunato2018science}
Efforts on quantifying academic performance at individual and group levels use productivity measures such as number of publications and impact indicators like citation counts.\cite{sinatra2016quantifying,wu2019large} 
Impact of academic mentorship and institutional quality for academic growth have been recently studied by using these bibliographic indicators.\cite{sekara2018chaperone,way2016gender,way2019productivity,ma2020mentorship}

We want to investigate academic productivity by utilizing the social aspect of doctoral studies.
We investigated the publication records of students obtained from an online service, called Dimensions, for 2824 former doctoral students.\cite{pub.1106289502} 
By conducting a regression analysis, we analyzed the role of academic support network by considering number of mentions and sentiment scores of acknowledgement to estimate publication count during the doctoral studies while controlling for disciplinary differences and gender.
We employed an Inverse Gaussian regression model to estimate the parameters and their significance, since the target variable is the publication count and it approximately follows an Inverse Gaussian distribution (see SI:Sec-\ref{secSI:regression_analysis}).
To capture the productivity during doctoral studies, we used number of publications as a measure and considered the period of doctoral studies four years before graduation and four years after to account for the work in submission or in progress during the thesis defense. Results of the regression analysis are summarized in Table~\ref{tab:regression}. By analyzing the regression coefficients and the significant variables, we assessed the factors influencing the academic productivity and the social determinants of the doctoral students performance.

\begin{table}
\caption{\textbf{Regression results.} Inverse Gaussian model for explaining productivity by gender, discipline and textual features extracted from dissertation text.}
\centering
\renewcommand{\arraystretch}{0.6}
\begin{tabular}{ p{4cm} c  c  c }
 \hline
 \hline
 \centering
  &  & Publication Count &  \\\cmidrule{2-4}%
  & \emph{All Students}& \emph{Female Students} & \emph{Male Students} \\
 \hline
 gender & -0.1839*** &  & \\
 & (0.038) &  & \\
 \hline
 life\_earth & -0.1218* & -0.0943 & -0.1544\\
 & (0.058) & (0.086) & (0.079)\\
 math\_comp & -0.0032 & 0.0420 & -0.0318\\
 & (0.063) & (0.123) & (0.077)\\
 phys\_eng & 0.2468*** & 0.1171 & 0.2623***\\
 & (0.057) & (0.104) & (0.069)\\
 soc\_hum & -0.3968*** & -0.3907*** & -0.4404***\\
 & (0.048) & (0.066) & (0.073)\\
 \hline
 family\_sentiment & -0.0375 & -0.0426 & -0.0566\\
 & (0.047) & (0.073) & (0.061)\\
 \hline
 academic\_count & 0.0044* & -0.0011 & 0.0071*\\
 & (0.002) & (0.003) & (0.003)\\
 family\_count & 0.0004 & 0.0044 & -0.0027\\
 & (0.004) & (0.006) & (0.005)\\
 friends\&colleagues\_count & 0.0037 & 0.0029 & 0.0073\\
 & (0.003) & (0.005) & (0.005)\\
 administration\_count & 0.0018 & -0.0024 & 0.0048\\
 & (0.002) & (0.003) & (0.003)\\
 spiritual\_count & 0.0032 & -0.0506 & 0.0083\\
 & (0.210) & (0.040) & (0.025)\\
 \hline
 Constant & 2.1818*** & 2.0914*** & 2.2291***\\
 & (0.210) & (0.329) & (0.276)\\
 \hline
 Observations & 2,824 & 1,099 & 1,725 \\
 AIC Score & 16,206 & 5,831 & 10,383 \\
 \hline
 \multicolumn{4}{@{}l}{\footnotesize Standard errors in parentheses \hspace{1cm} $^{*} (p < 0.05)$; $^{**} (p < 0.01)$; $^{***} (p < 0.001)$}
 \footnotesize 
\end{tabular}
\label{tab:regression}
\end{table}

We investigated the regression analysis to validate our earlier observations about the gender and disciplinary differences.
It was shown in the literature on research outcomes that women have slightly less publication rates than men while the difference can be attributed to various systematic biases in academia.\cite{doi:10.1177/0306312705052359,10.2307/4106103,lariviere2013bibliometrics} 
Especially for STEM fields, empirical data reveals considerable gender variations in number of citations, publication counts and the impact of their academic careers.\cite{Huang4609,AbramoGiovanni2009Gdir} 
This phenomenon can be explained by several factors; it is possible to consider that
women are underrepresented in scientific cooperation and publishing and struggle with implicit biases since they are more likely to play a significant role in parenting,\cite{kyvik1996child} obtain less institutional assistance and have more service duties,\cite{duch2012possible} or the systematic undervaluation of women’s involvement and their invisibility in scientific research, known as the ``Matilda Effect.''\cite{rossiter1993matthew} 
Consistent with the literature, our model have demonstrated that female productivity is lower than that of males when considering simply the number of publications (M = -0.1839, 95\% CI [-0.258, -0.110]).
These gender differences imply that studying the doctoral process may help to better understand the above mentioned adverse conditions. 

Another important aspect explaining the productivity is the academic discipline because publication counts vary from one field to another,\cite{sabharwal2013comparing} which is a key indicator of quality in higher education since the research performance has an influence on rewards, tenure, promotion decisions and staff recruitment.\cite{costas2010bibliometric, bland2006impact, blackburn1995faculty} 
Therefore, it is essential to demonstrate and explain the alterations between scientific fields. When Biology \& Health Sciences is taken as the reference group, our model indicates that while the Physics \& Engineering students are associated with more publications (M = 0.2468, 95\% CI [0.136, 0.358]), Life \& Earth Sciences (M = -0.1218, 95\% CI [-0.235, -0.009]) and Social Sciences \& Humanities (M = -0.3968, 95\% CI [-0.492, -0.302]) students are affiliated with less number of papers, as also suggested in the Becher's work on disciplinary differences.\cite{becher1994significance} However, when we controlled for the gender variable, our findings showed that only being a Social Sciences \& Humanities (M = -0.3907, 95\% CI [-0.520, -0.262]) student is negatively correlated with publication counts for females. On the other hand, male Physics \& Engineering (M = 0.2623, 95\% CI [0.126, 0.398]) students are associated with more publications while Social Sciences \& Humanities (M = -0.4404, 95\% CI [-0.584, -0.297]) are associated with less. These results might indicate the under representation of female doctoral students in Physics \& Engineering fields.

Aside from the demographic aspects, our results demonstrated that the number of people from academic network mentioned in acknowledgements is associated with more publications (M = 0.0044, 95\% CI [0.000,  0.009]). However, when controlled for genders, the regression analysis suggested that this statement holds only for male students (M = 0.0071, 95\% CI [0.002, 0.013]).

\begin{figure}[htp]
    \centering
    \includegraphics[width=1\linewidth]{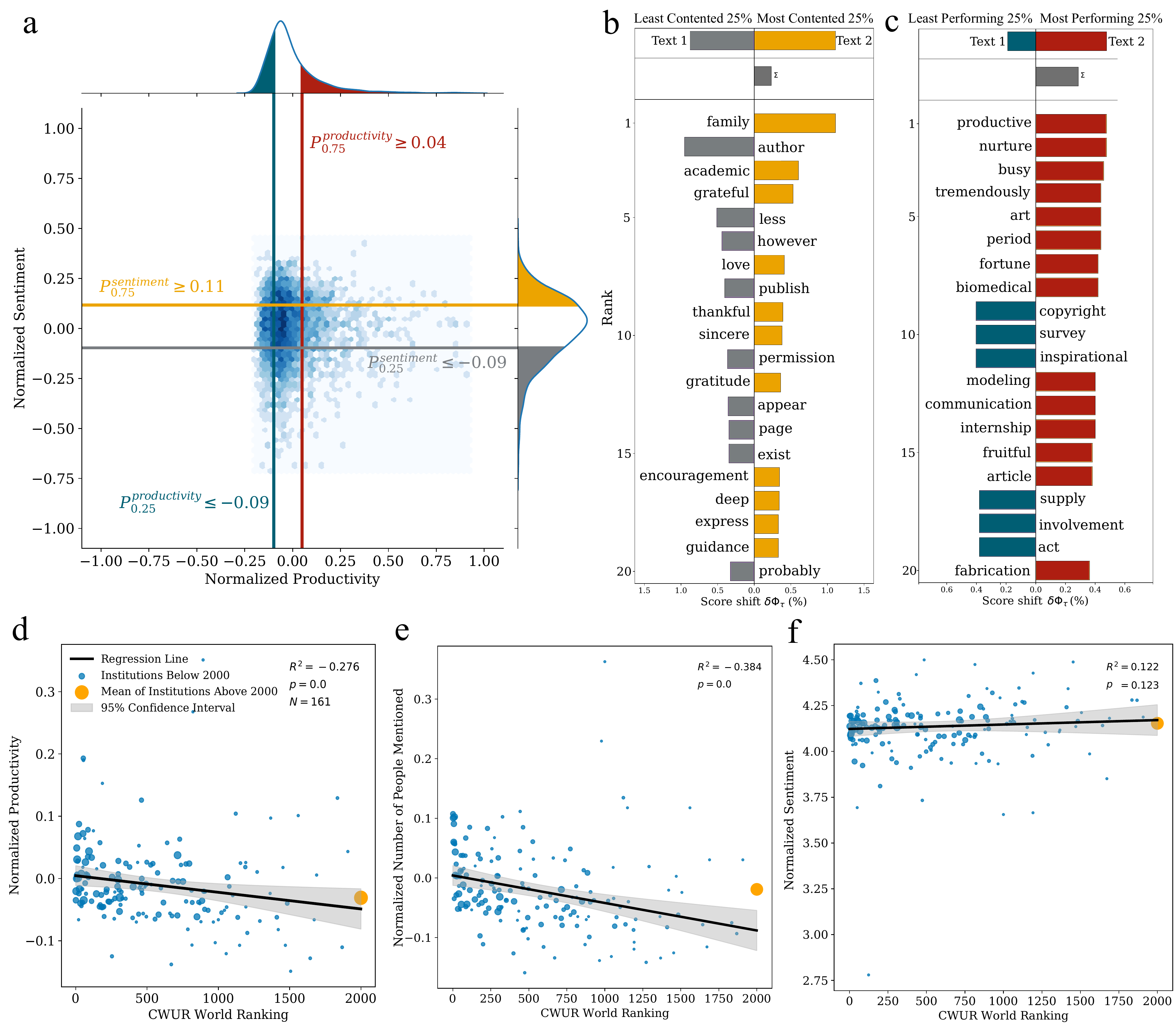}
    \caption{\textbf{Determinants of academic productivity and linguistic differences between extreme cases} Sentiment and productivity levels of students were normalized based on their disciplines and genders (a). Word usage differences are quantified by JS Divergence scores and compared students at the first and third quartiles based on normalized sentiment and productivity (b, c). Relationship between CWUR World rankings and normalized productivity levels (d). Relationship between CWUR World rankings and normalized number of mentions in acknowledgements (e). Relationship between CWUR World rankings and normalized sentiment scores in acknowledgements (f). R-squared values denote Spearman's rank correlation. Size of blue dots is proportional to the number of theses from these institutions.}
    \label{fig:sentiment-productivity}
\end{figure}

Our models do not suggest a statistically significant relationship between the rest of the variables and the number of publications; however revealed the influence of gender and discipline on productivity.
Therefore, we normalized sentiment scores and publication counts between zero and one at the individual level by taking into account gender and discipline of a student.
More clearly, we filtered out each gender-discipline pairs from our data and normalized publication counts by min-max scaling. These values are then subtracted from the group mean to center around zero.
Distributions of normalized sentiment and productivity scores are shown in Fig.\ref{fig:sentiment-productivity}(a).

Empirical and visual evidence shows no sign of significant links between sentiment and productivity levels. Additionally, we compared the language characteristics of extreme cases for both productivity and sentiment levels to help us understand the mindsets of people from upper and lower quantile of the distributions. To achieve this, we inspect the word usage differences in two groups by using Jensen-Shannon (JS) divergence for words that are used more than 10 times in each group. These words are then represented as word-shift graphs as shown in Fig.\ref{fig:sentiment-productivity}(b) for sentiment and Fig.\ref{fig:sentiment-productivity}(c) for productivity.\cite{gallagher2021generalized}

Sentiment scores depend on content and context of texts. Hence, there expected to be certain alterations between relatively more positive and negative acknowledgements. As seen on Fig.\ref{fig:sentiment-productivity}(b), we observed that most contented 25\% of PhD students emphasize gratitude by giving more space in their narratives to words such as \textit{grateful}, \textit{gratitude}, and \textit{thankful}.
These results also conformed to the past work which suggests that expressing gratitude helps to increase well-being.\cite{emmonse2003counting,killen2015using}
Our results also demonstrate that both family and the academic environment are more frequently mentioned in the narrative of the most contented 25\%.
Fig.\ref{fig:sentiment-productivity}(c) illustrates the JS divergences of words across most and least performing doctoral students. It is apparent that those who over-perform their counterparts emphasize more endeavour related concepts such as \textit{productive}, \textit{busy}, \textit{internship}, and \textit{article}.

\textbf{Institutional ranking and student performance}
Since the most well-known university ranking organizations such as Quacquarelli Symonds (QS), Times Higher Education (THE), and Center for World University Rankings (CWUR) employ ``number of research papers published'' as a factor in their ranking, we assumed that productivity levels of doctoral students may have associations with the success of their institutions. We present analysis on CWUR ranking, since it provides a more granular and longer list, but our results are consistent for other ranking systems (see SI:Sec-\ref{secSI:uni_rankings}).

We investigated the relationship between university rankings and productivity of graduate students. 
We found that university rankings have significant positive correlation with the number of publications (Fig.\ref{fig:sentiment-productivity}(d)).
Research environments in these institutes provide more opportunities to publish and introduce them a broader collaboration network as well, partially observed by total number of people mentioned (Fig.\ref{fig:sentiment-productivity}(e)). 
Number of people mentioned in dissertations have a higher correlation with institute ranking than the productivity levels, suggesting environment cultivate institutional success more than publications alone.
However, there is no associations between sentiment of a doctoral student with respect to the ranking of their institutions (Fig.\ref{fig:sentiment-productivity}(f)) meaning that the top-ranked institutes provide advantage in professional growth while well-being of the doctoral students mostly determined by their academic support networks.

\newpage

\section*{Discussion}

Our research uncovered the network of support providers, assisting doctoral students in achieving their goals. We showed that there exist gender and disciplinary differences in acknowledging support providers and sentiment scores when mentioning different communities. Since  acknowledgements often appear as the sole section in which students talk about their experience as doctoral candidates, it is noteworthy to observe that the link between productivity levels and their academic support networks can be revealed.

Our results showed that the number of publications among doctoral students varies by academic discipline, with students in social sciences and humanities publishing the least and students in physics and engineering publishing the most. 
Our data also suggested that productivity is positively correlated with the number of people mentioned from academic environment when publication counts are normalized with regard to gender and discipline. 
We showed that female students are more likely to acknowledge each support provider group in a more positive sentiment. They did, however, mentioned fewer people from their workplace and published fewer academic publications, implying that women’s professional support networks are limited, leading to lower productivity levels at the time they receive their doctoral degrees.
Our results also demonstrated that schools with higher rankings provide PhD students wider networks, in which academic environment is significantly bigger and productivity levels as a result are higher. In fact, it is shown in the literature that as the number of writers rises, so does the impact of the research,\cite{lariviere2015team} highlighting the importance of young scholars' academic networks.

Quantitative analysis of acknowledgement texts provided a deeper insight on social interactions and experiences of doctoral students as well.
Our results suggested that the narrative of the most performing 25\% is more centralized on endeavour-related content compared to the least performing 25\%. Similarly, most contented 25\% show more gratitude towards their family and academic environment compared to the least contented 25\%.

It is crucial to note that while support providers from academic communities have a positive influence on productivity, overall well-being of a student require contributions from social interactions with family and friends and administrative support from their institute.
Our results showed that higher university rankings or productivity levels do not lead to a higher sentiment reflected towards doctoral experience, but positive influence in their professional growth.

Analyzing thousands of acknowledgement sections, we created an alternative angle reflecting social aspects of the doctoral studies where friends, families, colleagues, and administrative staff have different roles to play ensuring utmost performance and well-being of the student.
Therefore, instead of directly analyzing publication counts or number of citations to explain doctoral studies, it may be better to embrace a new approach where students' well-beings and academic support networks are also put forward.
People compare themselves to those who are similar to them with regard to demographic and social proximity and how individuals evaluate their own subjective well-being and happiness depends on those of others.\cite{posel2011relative, de2012relative} 
It is also known that ``success narratives'' have an impact on the reader's judgements and decisions,\cite{lifchits2021success} which may imply that when doctoral students compare themselves with their counterparts, it would decrease their subjective well-being.
Future work may contribute to a profound understanding of how support networks influence productivity in late career stages and researchers' overall well-being by reaching out to people and possibly conducting a survey. It is also important to collect theses published all around the world to improve the representativeness of the data and observe how cultural aspects influence the way of doctoral students acknowledge their support providers.

\section*{Methods}
\textbf{Data collection and information extraction} To run a large-scale analysis of dissertation acknowledgements, we retrieved data from \texttt{pqdtopen.proquest.com}, also called PQDT-Open, which provides a publicly accesible thesis dissertation data that allows our work to be reproducible. We collected the dissertations by scraping the data directly from the website using Selenium library offered in Python.
We have gathered documents for 47,000 researchers, and 26,264 of them are doctoral dissertations, written between 2000 and 2020. This collection of dissertation data also included metadata on dissertation abstract, title, author name, university, year of publication, page number, advisor name, department, subjects, and keywords. 
To extract the acknowledgement subsection from these dissertations, we parsed raw data obtained in Portable Document Format (PDF). We used PyMuPDF library to extract textual information for each page, and then, we utilized a rule-based approach to identify pages that are likely to belong to acknowledgement subsection – e.g., accepting the pages with the first word being ``Acknowledgements,'' or ignoring pages that contain text such as ``Table of Contents,'' ``List of Figures,'' or ``Appendix.''

\textbf{Data enrichment} Although there is a rich metadata provided by PQDT-Open, there was no gender information or discipline category given. 
For the former one, we inferred students’ genders using their first names. We used the online service, called \texttt{genderize.io}, which relies on a database that has over 110 million entries from 242 countries to examine whether a name is more frequently used amongst females or males.
For the latter one, although dissertation subjects were given in the metadata, it was not feasible to run an analysis uncovering the disciplinary differences since we identified 572 unique subjects listed in total. Hence, to have a clearer view of disciplinary differences, we divided the subjects into 5 categories.\cite{lamers2021measuring} Although there is no categorization agreed in the literature and guideline or consensus on how research fields should be classified, it could be done by following previous research efforts. To assign each subject into one category, two authors separately labeled each subject with a discipline and reached to an 85\% agreement and $0.75$ Cohen's Kappa\cite{cohen1960coefficient} score for inter-annotator reliability. A detailed list of category – subject is given under SI:Sec-\ref{secSI:discipline_categories}

\textbf{Bootstrapped estimates of sentiment and counts}
The bootstrap method is a statistical methodology that involves averaging estimates from several small data samples to infer statistics about a population.
We employed this approach to estimate mean and confidence interval of sentiment scores and number of mentions to be able to observe disciplinary and gender differences.
We repeated this procedure 5000 times for each estimation. We randomly drew 5\% of the population with replacement and calculated the mean of the sample. Using these mean values, we ran two-tailed T-tests to test significance of sample means ($p \leq 0.05$, $p \leq 0.01$, $p \leq 0.001$ and thresholds are denoted by ``*'', ``**'' and ``***'', respectively).

\textbf{Regression analysis}
Since linearity and normality assumptions do not hold in our case and our target variable (publication count of doctoral students) follows approximately an Inverse Gaussian distribution, we employed a generalized linear model with Inverse Gaussian distribution.
After selecting the appropriate regression model, to detect multicollinearity and select the variables that are going to be used in regression analysis, we checked the variation inflation factor (VIF) and removed those which had higher than 10. The remaining variables were used in the regression analysis.

\section*{Acknowledgements} We would like to thank VRL Lab members, Qing Ke, and Nur Mustafaoglu for their feedback and fruitful discussions. O.V. also thanks his academic support network challenging him to do his best.

\newpage
\section*{References}
\bibliography{main}
\bibliographystyle{naturemag}

\newpage
% This will append SI material at the end of the text
\renewcommand{\thefigure}{SI-\arabic{figure}}
\setcounter{figure}{0}

\renewcommand{\thetable}{SI-\arabic{table}}
\setcounter{table}{0}

\setcounter{page}{1}

\section*{Supplementary Information}
\begin{figure}[htp]
    \centering
    \includegraphics[width=\linewidth]{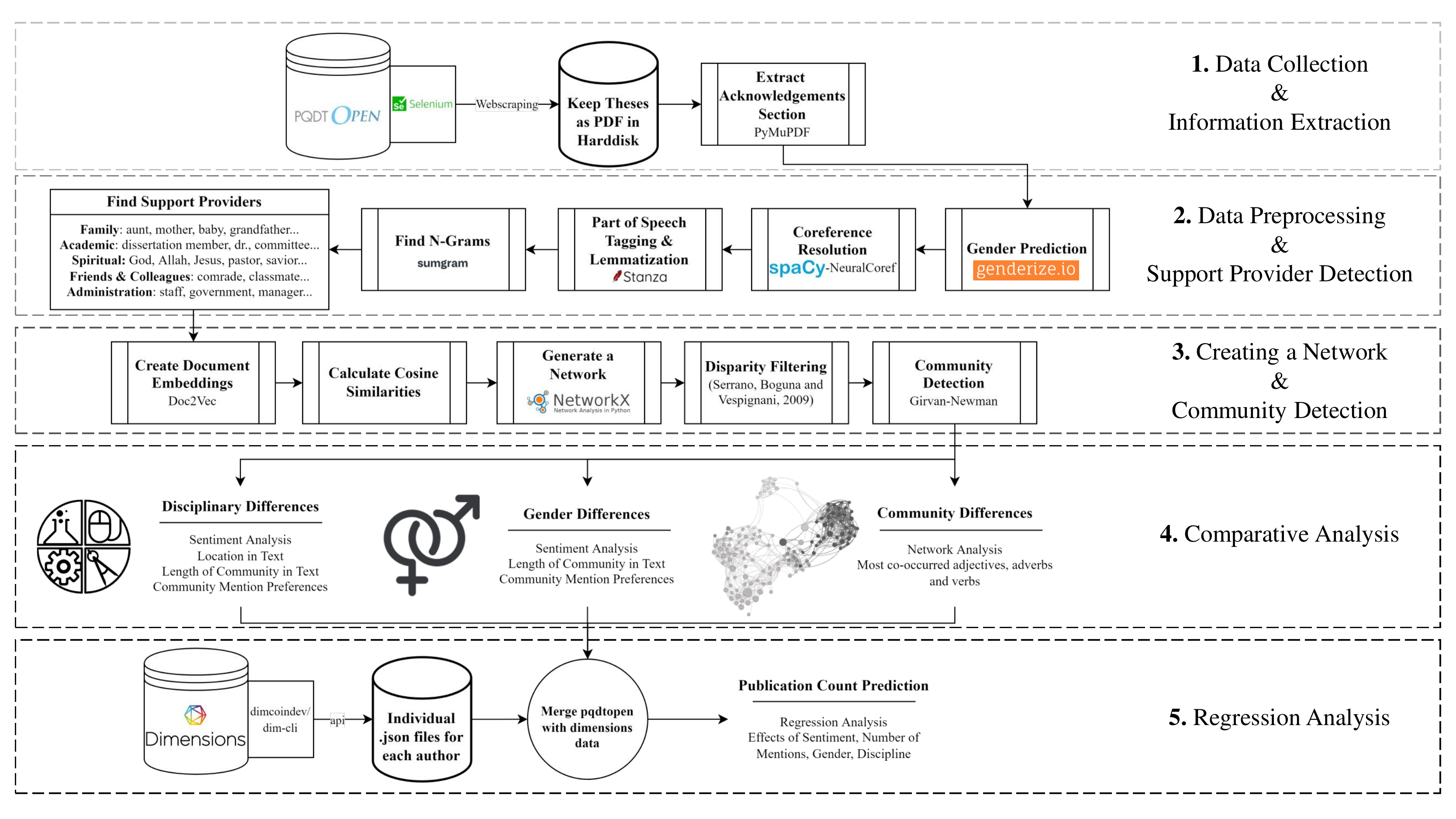}
    \caption{The Complete Flowchart of the Study}
    \label{figSI:flowchart}
\end{figure}

This section presents methodological step followed in this study. Fig.\ref{figSI:flowchart} demonstrates the overall framework. 

\setlength{\parindent}{0ex}

\newpage
\section{Data Collection and Information Extraction}
\label{secSI:datacollection}

To run a large-scale analysis of dissertation acknowledgements, we first needed a comprehensive and reliable database that captures metadata such as publication year, institution, thesis subject, and content of the study. Secondly, we had to extract acknowledgement subsection of theses and converted into a machine readable format. 

\textbf{Dissertation data scraping} We retrieved data from \texttt{pqdtopen.proquest.com}, also called PQDT-Open, which offers open access to dissertations from institutions all around the world, most frequently located in the United States. Since PQDT-Open provides a publicly accessible data that allows our work to be reproducible, we chose to collect the dissertations by scraping the data directly from the website using Selenium library offered in Python.

We have gathered documents for 47,000 researchers, and 26,264 of them are doctoral dissertations, written between 2000 and 2020. This collection of dissertation data also included metadata on dissertation abstract, title, author name, university, year of publication, page number, advisor name, department, subjects, and keywords. 

\textbf{Extracting acknowledgement text} To extract the acknowledgement subsection from the dissertations, we parsed raw data obtained in Portable Document Format (PDF). We used PyMuPDF library to extract textual information for each page, and then, we utilized a rule-based approach to identify pages that are likely to belong to acknowledgement subsection – e.g.,accepting the pages with the first word being ``Acknowledgements,'' or ignoring pages that contain text such as ``Table of Contents,'' ``List of Figures,'' or ``Appendix.'' When we encountered a page that starts with ``Acknowledgements,'' we took the whole text from that page. If text in a page exceeds 1100 characters - which was the observed mean character limit for one page in our case, our system checked the next page if there is any sign of that page not belonging to an acknowledgement section.

\newpage
\section{Data Preprocessing} \label{secSI:dataprocessing}
Our analysis relied on individual and institutional offices supporting a doctoral student. We analyzed acknowledgement text to determine these entities and extract information on how they are referred and acknowledged. We then constructed an ``academic support network'' based on their contextual similarities.

\subsection{Preprocessing the Text and Metadata \& Support Provider Detection}\label{SI:preprocessing_the_text}

In the acknowledgement section, support providers are introduced by their names and affiliation, but they can also be referenced in a series of consecutive sentences by personal pronouns. We use Coreference Resolution to identify pairs of references that are associated with the same entity. Coreference resolution approach helped us to link the support providers with all other related words and was the first step in the text preprocessing.

Next, we lemmatized and applied part-of-speech tagging (POS tagging) by using Stanza library to create different functional words. Since some POS types such as punctuation, numerals, and conjunctions are not relevant to our analysis, we built our models by using only the words that are tagged as adjective, adverb, verb or noun.

In the last step, we employed Sumgram library to identify n-grams that are used regularly in documents, therefore helping us to find some of the word groups that might indicate a support provider, such as ``thesis advisor.''

By using the corpus that we created after preprocessing the data, we employed both manual and data-driven approaches to find the support providing entities within text. Firstly, we created a corpus of word and n-grams that have been found in acknowledgements, we went over the word or word groups that have more than 50 occurrences by manual inspection to clarify whether these tokens belong to a support provider such as mother, girlfriend, father, colleague, thesis advisor, God etc.

To complement with the previous approach, we used \textit{Word2Vec} to identify the remaining support providers. Once we created a list of support providers from manual inspection, we examined 10 most similar words for each support provider and included them to the collection if needed. This hybrid process helped us to minimize any bias when determining the words and phrases that are being examined in this study.
Among 155 words support providers identified in this analysis, we removed the ones that occur less than 50 times in acknowledgements, leaving us 144 support providers which can be further examined under Table~\ref{table:list_of_support_providers}.

\subsection{Creating a Network \& Community Detection}
\label{secSI:creating_a_network} 
We created word embeddings for support providers by using Doc2Vec model trained on the dissertation acknowledgement corpus. Each sentence that a support provider mentioned is tagged with the related support provider and these tagged document vectors are given as inputs to the model. We used the default parameters in the model, which provides 100 dimensional vectors for each support provider. These vectors are then used as feature vectors for support providers.

After learning the vector representations for support providers, we created a network representation where nodes correspond to support providers and their sizes are determined by their number of occurrences. The edge weight between each pair of nodes is defined as the associated cosine similarity between vectors representations learned from. Doc2Vec model.

\begin{figure}[htp]
    \centering
    \includegraphics[width=9cm]{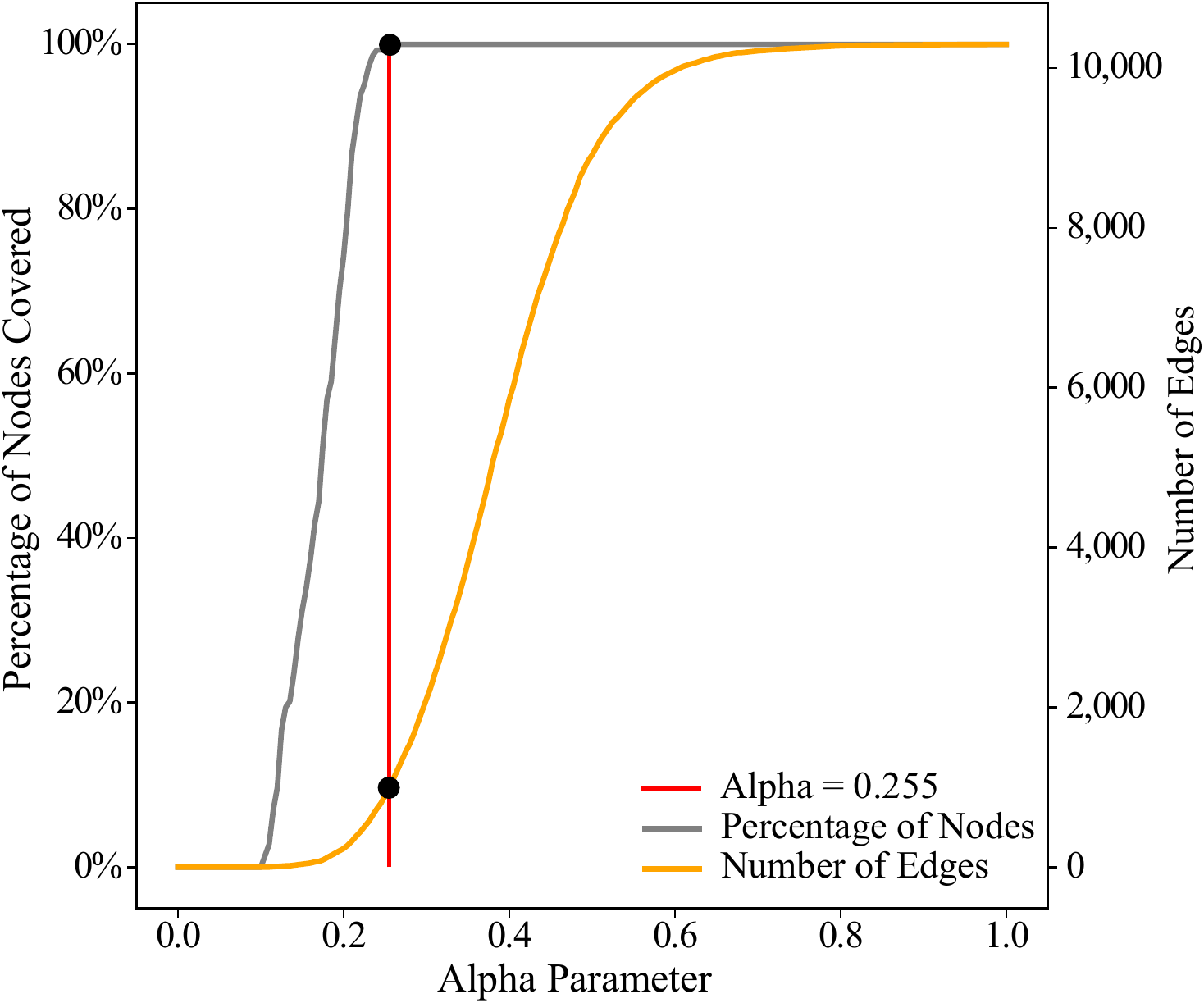}
    \caption{Percentage of Nodes Covered and Number of Edges for Associated Alpha Values in Disparity Filtering}
    \label{fig:disparity_filtering}
\end{figure}

Support providers network will be fully connected since edge weights are computed using cosine similarities between dense vector representations. We filtered edges using disparity filtering \cite{serrano2009extracting} to focus on statistically significant associations. Resulting network gave us the sparsest network possible with 144 nodes and 759 edges as shown in Fig.\ref{fig:disparity_filtering}. Subsequently, to detect the communities in the network, we used Girvan-Newman algorithm \cite{girvan2002community} and observed the formation of the 5 communities as given under Table\ref{table:list_of_support_providers}.

\begin{figure}[t!]
    \centering
    \includegraphics[width=17cm]{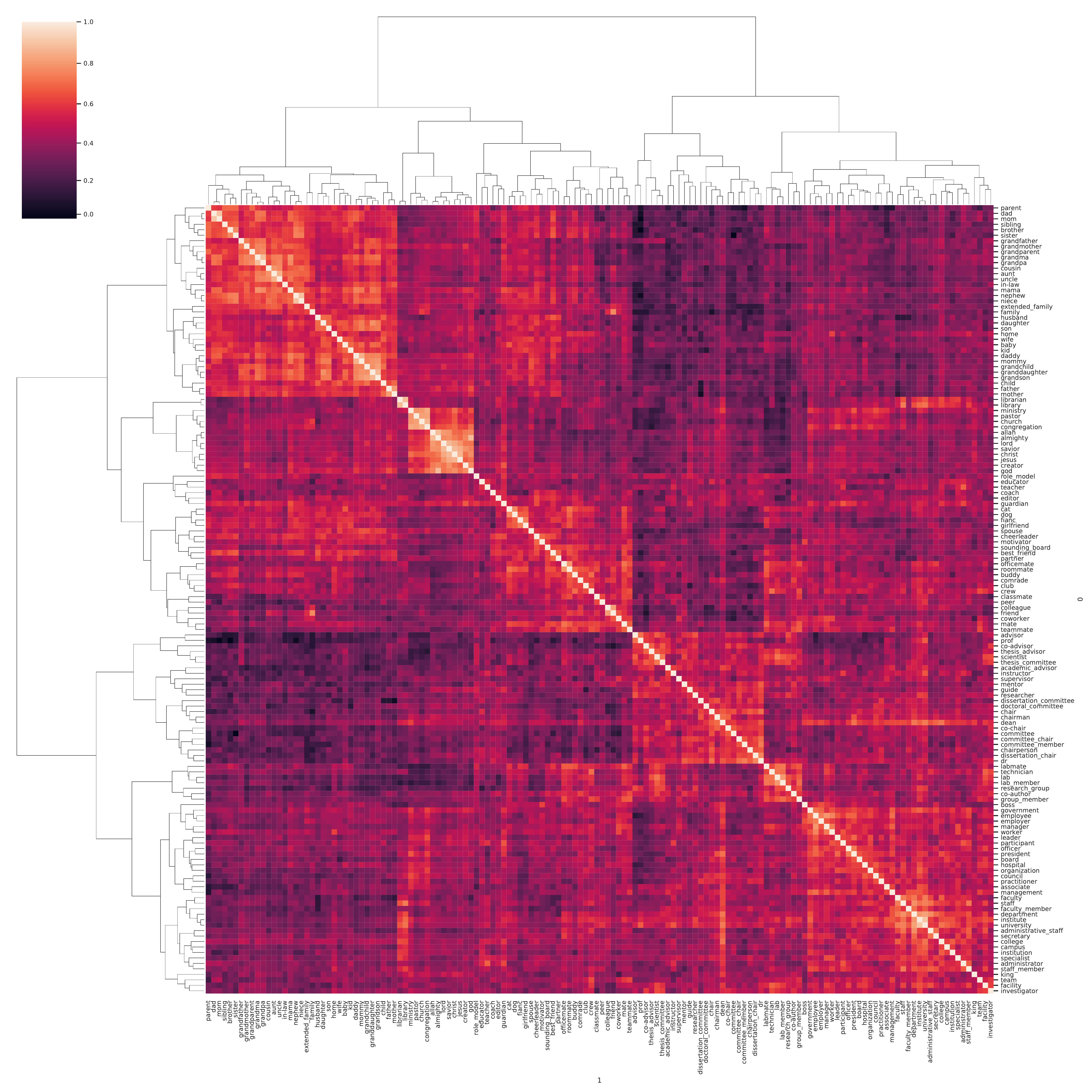}
    \caption{Support providers vector similarities using hierarchical clustering with complete linkage and dendograms computed following agglomerative approach.}
    \label{figSI:hierarchical-clustering}
\end{figure}

To check the validity of the communities, we performed a similar analysis using hierarchical clustering and obtained the results as shown in the Fig~\ref{figSI:hierarchical-clustering}, which provides 5 clusters that are significantly overlapping with the results obtained using network clustering approach.

\subsection{Methodological Toolkit}
\label{secSI:methodology} In this section, we will briefly introduce text analysis and natural language processing techniques.

\textit{Tokenization, Part of Speech Tagging (POS) and Lemmatization} Tokenization is the process of identifying meaningful units of a sentence, such as words, numbers, or punctuation. This gives us the capability of labelling the words in a text to a specific part of speech, regarding their definition and context (e.g., verb, noun, preposition) which is called part of speech (POS) tagging. At last, after detecting each word in the sentence, we can take off their inflectional suffixes, and then only consider the dictionary form of the word, which is called the lemmatization. E.g., running, ran and run will be all transformed into ``run.'' This process helps us to reduce the number of words and to detect the entities that are associated with the support providers while keeping the integrity of the meaning in texts.
For these processes, we used the pretrained, neural network based natural language processing package Stanza \cite{qi2020stanza}, offering a state-of-the-art performance on various NLP tasks.

\textit{Sumgram}
Adjacent sequence of N number of words is called an N-gram. E.g., ``nice picture'' is a 2-gram – or bigram, and ``beautiful Saturday morning'' is called a 3-gram. To detect N-grams, traditional models use a Bayesian approach to predict the occurrence of a word, based on the occurrence of its N – 1 preceding words. For example, to predict the occurrence of ``picture'' after ``nice'' occurs, it calculates  $P(picture ~ | ~ nice)$=(number of times ``nice picture''  occurs)/(number of times ``nice''  occurs). In here, depending on the model, number of occurrences can be document frequency or term frequency within a document.

While N-grams can be useful for several different tasks, our goal is to catch the words that together represent meaningful noun phrases. To achieve this, we used Sumgram library, which seeks to conjoin small N-grams by a function that uses an expanding window over words, and with another function (that works with a set of rules using POS tags) tries not to split multi word proper nouns. Furthermore, Sumgram considers document frequency instead of term frequency, thus giving every acknowledgement single vote for a single term. This is beneficial for our case since the length of acknowledgements vary from dissertation to dissertation.

\textit{Coreference Resolution}
Finding and pairing all textual references that are associated with the same entity is called coreference resolution. Since we seek to pair all the support providing entities with the other words that are used in the same sentence, changing the personal nouns into entities themselves helps to find all the pairs, which in turn allows us to make use of the whole acknowledgement.
In traditional models, coreference resolution is done by manually designed scoring algorithms,\cite{bagga1998algorithms} which may cause a variation in performance across different languages and datasets.\cite{clark2016improving}
We employed the Neuralcoref module based on Spacy parser, which is easy to use and provides a data-driven Word2Vec based approach for coreference resolution, rather than using instructions given beforehand (such as pronoun ``her'' is feminine, then bind it always with feminine words as ``sister,'' or ``mother'').

\textit{Word Embeddings}
Word embeddings are vector representations of document vocabulary, and which are capable of keeping semantic and syntactic properties of words. The idea here is that words that occur in the same contexts tend to have similar meanings.\cite{harris1954distributional} Therefore, they allow us to find the similarities between words. 

We tried three different approaches to challenge the robustness of our results and decided to continue with Doc2Vec because of the following three reasons:

\begin{itemize}

\item[(1)] It is proven to be capable of capturing the context of documents – in this case, sentences that include support providers – and shown to be more successful than traditional methods such as bag-of-words.
\item[(2)] It is much more memory efficient than TF-IDF method. While all the support providers are represented as 100-dimensional vectors with the Doc2Vec method, TF-IDF uses the whole words in the corpus – even though we remove the words that have occurred once or twice in the dataset, there are 15 thousand dimensions left for each support provider.
\item[(3)] Since Doc2Vec is not pre-trained and learns the word representations directly from the dataset, its outputs are context-dependent and well-crafted for our task. Therefore, we did not use any pre-trained, neural network based models such as GloVe \cite{pennington2014glove}.
\end{itemize}

\textit{Sentiment Analysis}
We used a pretrained BERT model ~\cite{devlin2019bert} for sentiment analysis toolkit ``bert-base-multilingual-uncased-sentiment'' for this task.\footnote{\url{https://huggingface.co/nlptown/bert-base-multilingual-uncased-sentiment}}

When running analyses on acknowledgements, we considered each sentence separately, which gave us the opportunity to link the support providers with their associated sentiment scores. After forming the communities, we aggregated the sentiment scores of the support providers.

\textit{Bootstrapping Method}
We employed bootstrap sampling approach to estimate mean and confidence interval of sentiment scores over a population. We repeated this procedure 5000 times for each estimation. We have randomly drawn 5\% of the population with replacement and calculated the mean of the sample. Using these mean values, we ran two-tailed T-tests to see whether difference between two means is significant or not ($p \leq 0.05$, $p \leq 0.01$, $p \leq 0.001$ and thresholds are denoted by ``*'', ``**'' and ``***'', respectively).

\begin{table}[ht]
\small
\caption{List of support providing groups and words or phrases associated with these groups}
\begin{tabularx}{\textwidth}{|p{0.20\linewidth}|X|}
\hline
\textbf{Support providers} & \textbf{Keywords used for identification}\\
\hline
\hline
    Academic & 
    academic advisor, advisor, chair, chairman, chairperson, co-advisor, co-chair, committee, committee chair, committee member, dissertation chair, dissertation committee, doctoral committee, dr, guide, investigator, lab, mentor, prof, research group, researcher, role model, scientist, supervisor, thesis advisor, thesis committee \\
\hline
    Family & 
    aunt, baby, best friend, brother, child, cousin, dad, daddy, daughter, extended family, family, father, fiancé, girlfriend, grandchild, granddaughter, grandfather, grandma, grandmother, grandpa, grandparent, grandson, home, husband, in-law, kid, mama, mom, mommy, mother, nephew, niece, parent, partner, sibling, sister, son, spouse, uncle, wife, guardian \\
\hline
    Administration & 
    administrative staff, administrator, campus, college, council, dean, department, facility, faculty, faculty member, institute, institution, instructor, librarian, library, practitioner, secretary, staff, staff member, university, associate, king, educator, specialist, board, boss, coworker, employee, employer, government, hospital, leader, management, manager, officer, organization, president, worker, teacher, participant \\
\hline
    Friends \& Colleagues & buddy, cat, coach, dog, co-author, crew, group member, lab member, labmate, mate, officemate, team, teammate, technician, classmate, club, colleague, comrade, friend, peer, roommate, cheerleader, editor, motivator, sounding board \\
\hline
    Spiritual & 
    allah, almighty, christ, church, congregation, creator, god, jesus, lord, ministry, pastor, savior \\
\hline
\end{tabularx}
\label{table:list_of_support_providers}
\end{table}

\newpage
\section{Gender Inference \& Gender Based Differences} \label{secSI:gender_inference}

Since information about authors' gender is not provided in the metadata provided by PQDT-Open, we inferred gender information using first names of the authors. For this task, we used the online service named \texttt{genderize.io}. It relies on a database that has over 110 million entries from 242 countries to examine whether a name is more frequently used amongst females or males.

\begin{figure}[htp]
    \centering
    \includegraphics[width=16cm]{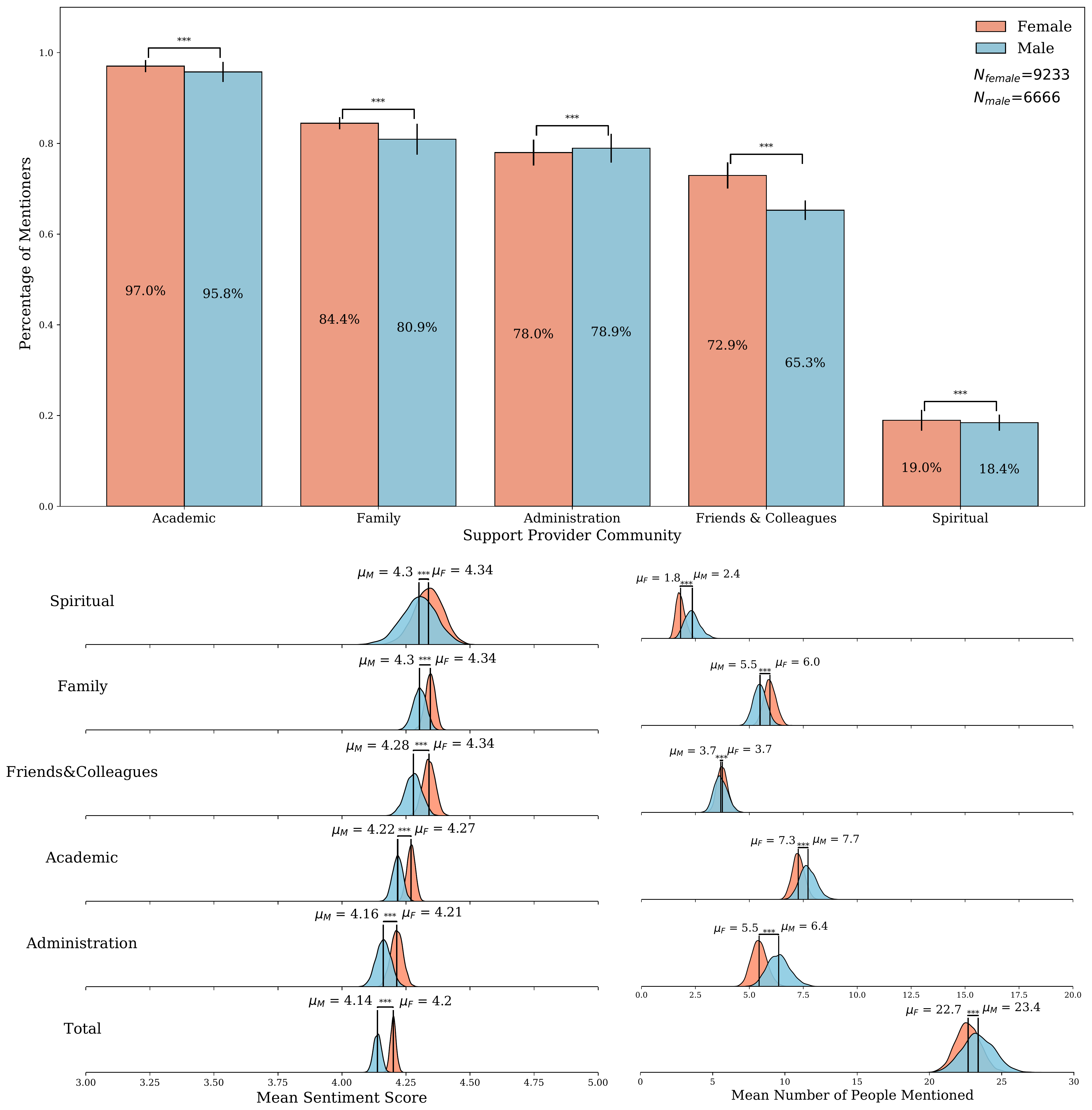}
    \caption{\textbf{Gender Based Differences in Terms of Communities for Social Sciences and Humanities Students}}
    \label{gender_based_soc_hum}
\end{figure}

\begin{figure}[htp]
    \centering
    \includegraphics[width=16cm]{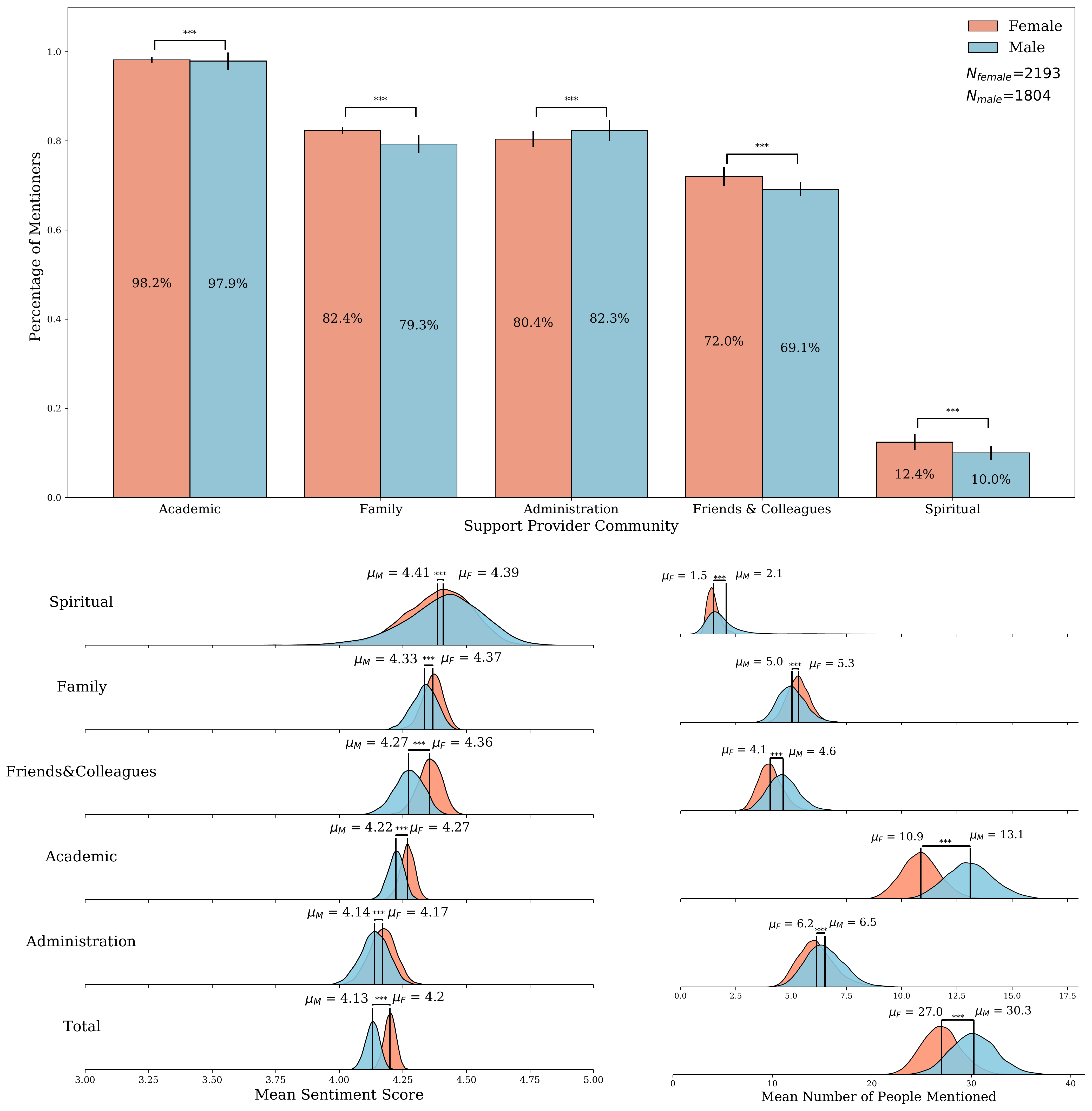}
    \caption{\textbf{Gender Based Differences in Terms of Communities for Biology and Health Sciences Students}}
    \label{gender_based_bio_health}
\end{figure}

\begin{figure}[htp]
    \centering
    \includegraphics[width=16cm]{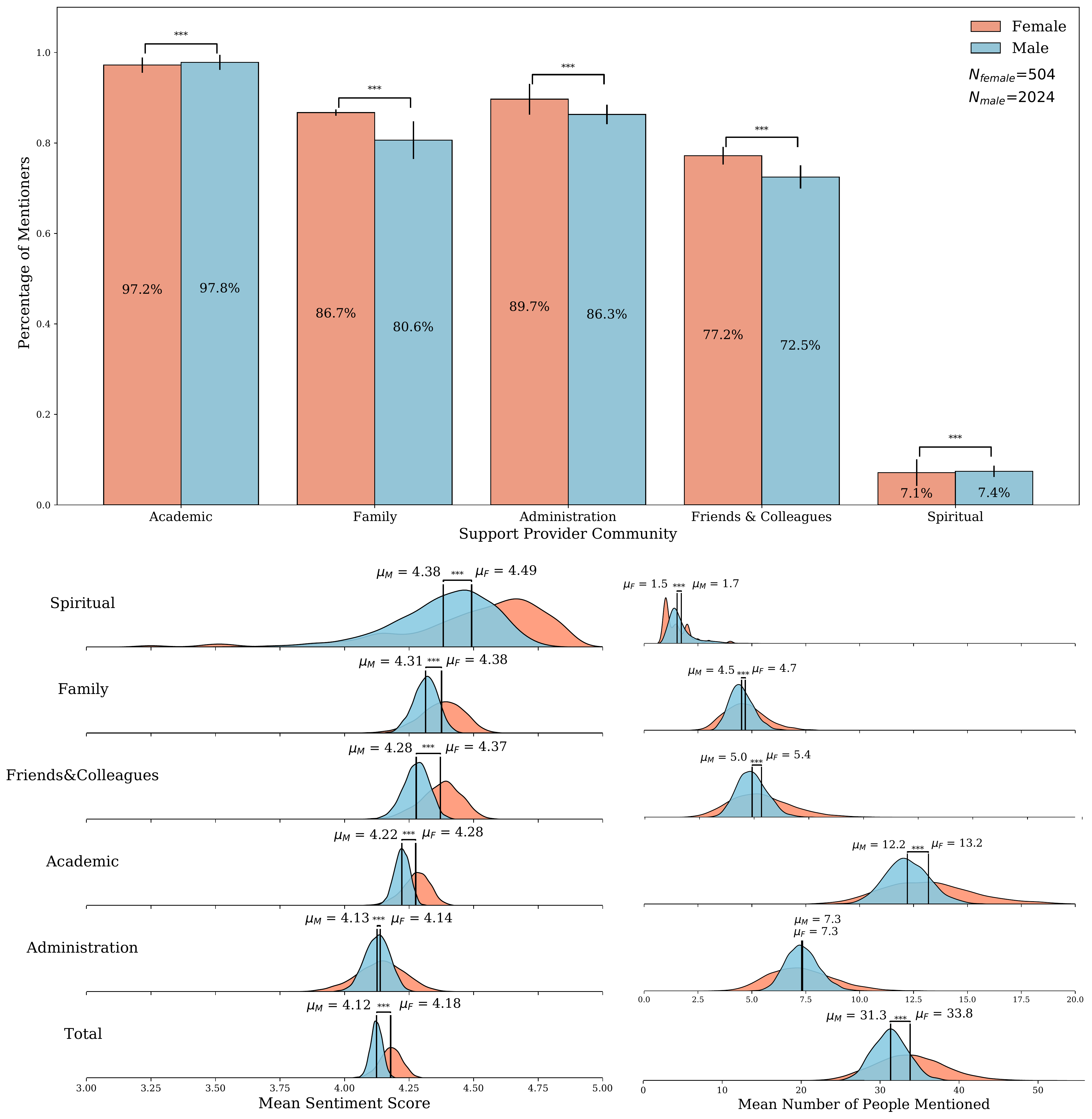}
    \caption{\textbf{Gender Based Differences in Terms of Communities for Physics and Engineering Sciences Students}}
    \label{gender_based_phys_eng}
\end{figure}

\begin{figure}[htp]
    \centering
    \includegraphics[width=16cm]{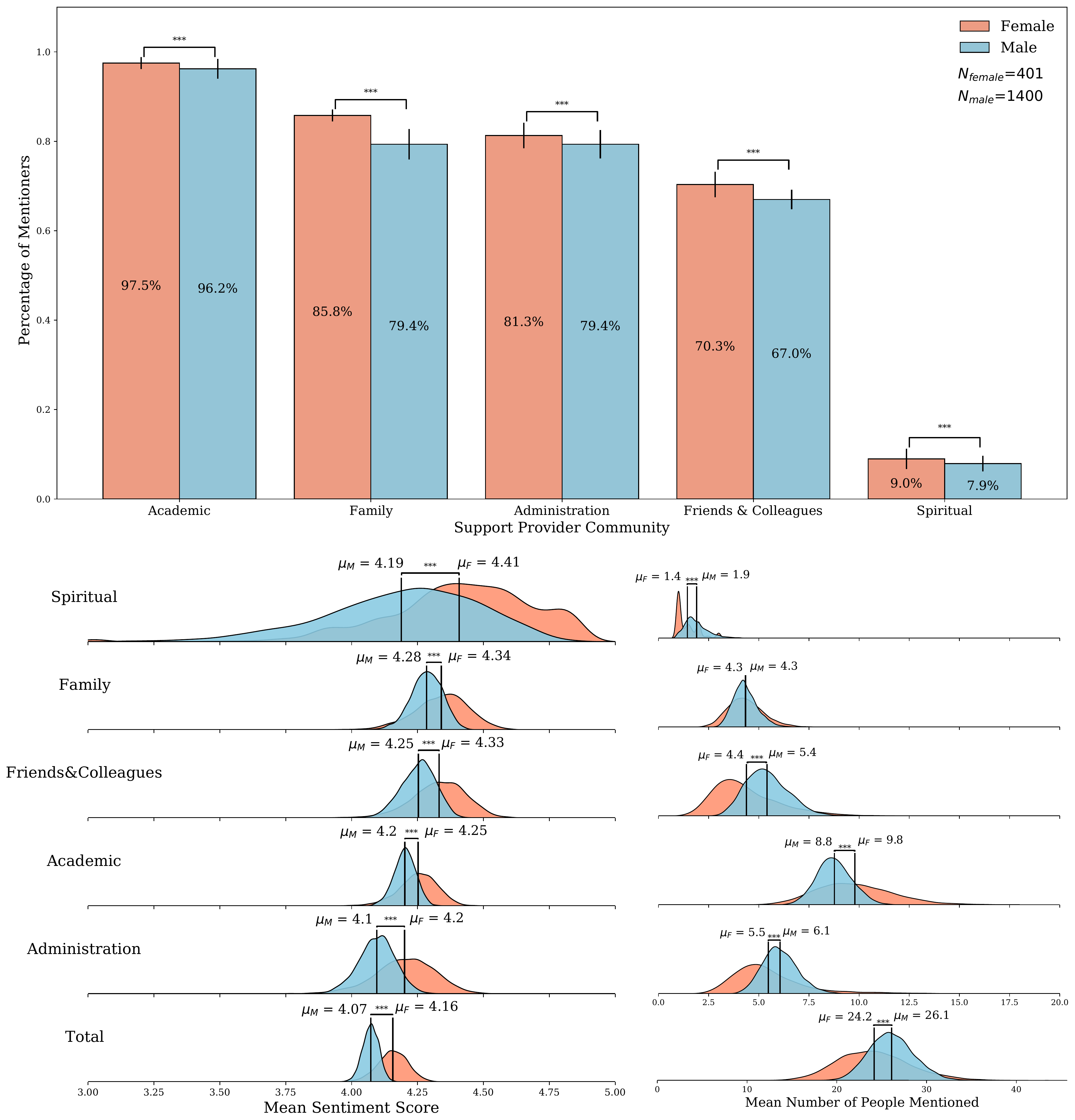}
    \caption{\textbf{Gender Based Differences in Terms of Communities for Mathematics and Computer Science Students}}
    \label{gender_based_math_comp}
\end{figure}

\begin{figure}[htp]
    \centering
    \includegraphics[width=16cm]{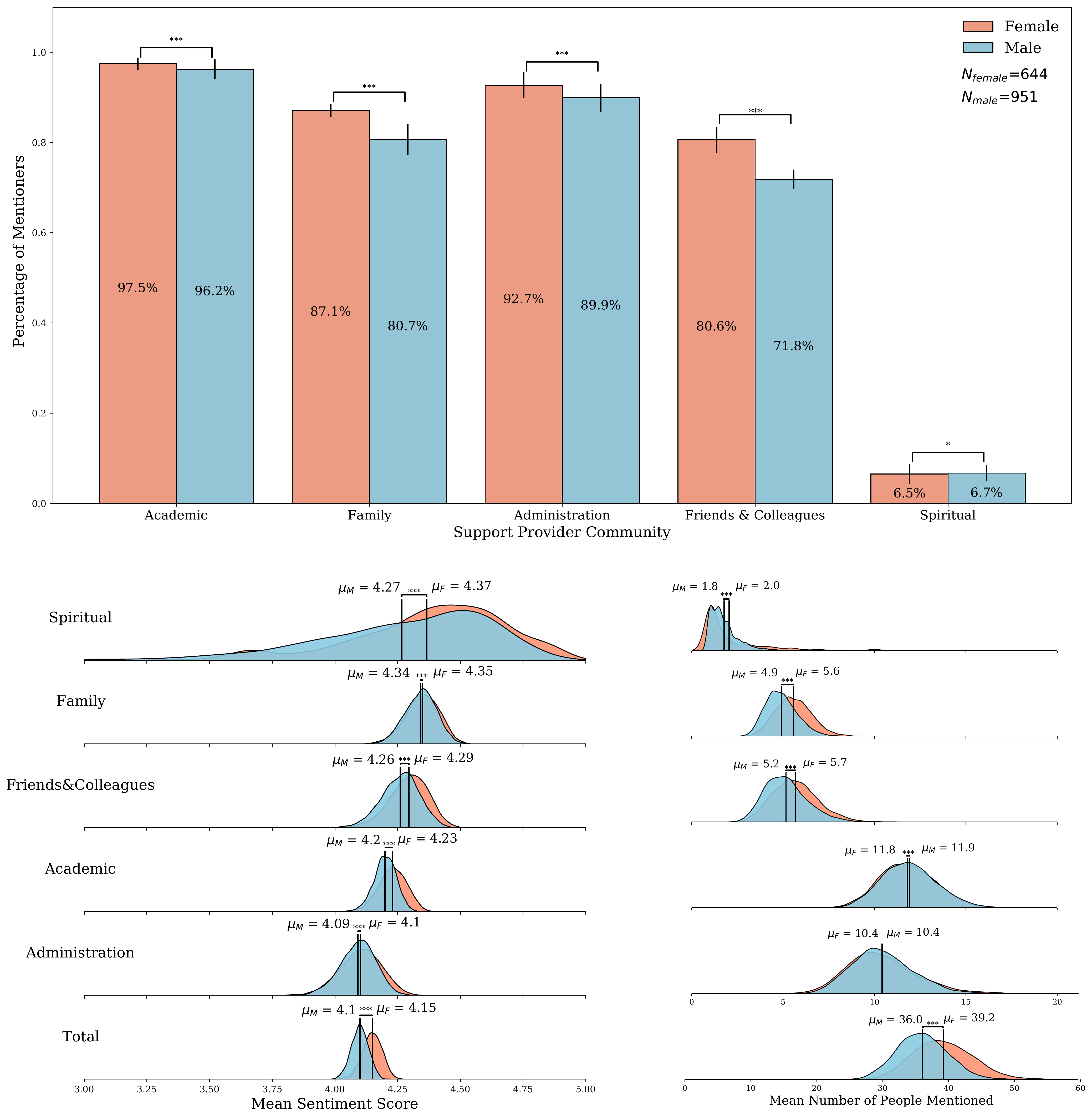}
    \caption{\textbf{Gender Based Differences in Terms of Communities for Life and Earth Sciences Students}}
    \label{gender_based_life_earth}
\end{figure}

We used this data to determine whether there are differences between females and males in terms of acknowledging the support providers as percentage of mentions, number of people acknowledged, and sentiment reflected towards a particular support provider group. Meanwhile the aggregated results for all disciplines are given in Fig.\ref{fig:genderdifferences}, we checked whether these results hold for disciplines as well. Our results are robust with some minor differences from one discipline to another (Figures in \ref{gender_based_soc_hum}, \ref{gender_based_bio_health}, \ref{gender_based_phys_eng}, \ref{gender_based_math_comp} and \ref{gender_based_life_earth}).

\newpage
\section{Determining Discipline Categories} \label{secSI:discipline_categories} 

Although there is no categorization agreed in the literature and guideline or consensus on how research fields should be classified, it could be done by following previous research efforts. Hence, to have a clearer view of disciplinary differences, we divided the subjects into 5 categories~\cite{lamers2021measuring}. To assign each subject into one category, two authors separately labeled each subject with a discipline and reached to an 85\% agreement and $0.75$ Cohen's Kappa\cite{cohen1960coefficient} score for inter-annotator reliability. Following categorization is used to define different disciplines throughout the paper.

{\footnotesize
\textbf{Biology \& Health Sciences}: nursing (991), public health (670), molecular biology (583), neurosciences (517), mental health (448), cellular biology (388), genetics (370), biomedical engineering (298), microbiology (257), medicine (239), epidemiology (225), biology (210), health sciences (209), oncology (194), physiology (189), immunology (185), organic chemistry (158), pharmacology (152), nutrition (134), evolution and development (125), virology (106), aging (102), developmental biology (82), surgery (82), kinesiology (76), pathology (74), physiological psychology (71), medical imaging (70), toxicology (63), biomechanics (62), pharmacy sciences (58), entomology (57), physical therapy (55), psychobiology (53), wildlife conservation (52), environmental health (49), animal sciences (48), alternative medicine (47), pharmaceutical sciences (45), parasitology (35), occupational health (34), food science (33), plant pathology (32), endocrinology (32), sexuality (31), audiology (30), systematic (26), veterinary services (26), obstetrics (26), ophthalmology (24), molecular chemistry (22), dentistry (22), medical personnel (22), molecular physics (21), organismal biology (20), systematic biology (16), neurology (15), histology (12), animal diseases (11), health care (10), biomedical research (10), pharmaceutical and medicine manufacturing (8), anatomy \& physiology (5), rehabilitation (5), general medical and surgical hospitals (4), health (3), osteopathic medicine (2), radiology (2), optometry (2), pharmaceuticals industry (1), research and development in the physical (1), nursing care facilities (skilled nursing facilities) (1), plant propagation (1), sports medicine (1), home health care services (1), radiation (1), pharmaceuticals (1)

\textbf{Life \& Earth Sciences}: ecology (429), biochemistry (428), environmental science (236), biophysics (228), chemistry (179), sustainability (165), geography (141), atmospheric sciences (120), geology (112), climate change (107), geophysics (96), hydrologic sciences (90), agriculture (87), plant sciences (84), environmental studies (81), geochemistry (71), environmental management (71), biological oceanography (67), natural resource management (66), zoology (66), water resource management (60), biogeochemistry (59), aquatic sciences (53), forestry (50), conservation (47), physical oceanography (45), plant biology (41), geographic information science (36), soil sciences (34), geomorphology (33), horticulture (30), meteorology (29), agronomy (29), atmospheric chemistry (27), physical geography (26), wildlife management (25), paleontology (25), limnology (23), land use planning (21), paleoclimate science (19), planetology (19), sedimentary geology (18), conservation biology (18), geotechnology (17), paleoecology (15), plate tectonics (14), chemical oceanography (14), botany (12), marine geology (12), water resources management (12), petroleum geology (10), mining engineering (10), geological (9), geophysical (8), wood sciences (8), macroecology (7), petrology (7), environmental geology (6), hydrology (6), aeronomy (6), mineralogy (5), atmosphere (5), petroleum production (4), geological engineering (4), urban forestry (3), geobiology (3), oceanography (3), geophysical engineering (2), and life sciences (1), polymers (1)

\textbf{Mathematics \& Computer Sciences}: computer science (1048), information technology (438), mathematics (355), information science (274), computer engineering (235), bioinformatics (233), applied mathematics (227), statistics (226), artificial intelligence (202), library science (79), biostatistics (66), computational physics (28), theoretical mathematics (27), computational chemistry (19), logic (9), systems design (6), information systems (6), software \& systems (1)

\textbf{Physics \& Engineering}:electrical engineering (576), mechanical engineering (406), materials science (374), engineering (313), physics (278), civil engineering (254), chemical engineering (205), optics (196), aerospace engineering (174), condensed matter physics (169), systems science (169), physical chemistry (156), analytical chemistry (143), astrophysics (137), urban planning (133), operations research (128), energy (127), astronomy (125), environmental engineering (121), nanotechnology (104), robotics (100), industrial engineering (100), particle physics (92), nanoscience (90), remote sensing (87), theoretical physics (86), inorganic chemistry (82), plasma physics (79), quantum physics (78), nuclear physics (63), transportation planning (60), electromagnetics (59), alternative energy (55), architecture (53), polymer chemistry (52), acoustics (48), mechanics (46), nuclear engineering (32), transportation (30), atomic physics (28), bioengineering (28), fluid mechanics (26), petroleum engineering (23), ocean engineering (22), agricultural engineering (22), automotive engineering (17), low temperature physics (16), applied physics (16), textile research (14), statistical physics (14), thermodynamics (10), naval engineering (9), landscape architecture (9), nuclear chemistry (8), high temperature physics (6), plastics (6), architectural engineering (6), architectural (5), robots (5), mining (3), fluid dynamics (3), gases (3), condensation (3), hydraulic engineering (3), automobile and light duty motor vehicle manufacturing (2), aerospace materials (2), mining and oil and gas field machinery manufacturing (1), molecules (1), all other miscellaneous manufacturing (1)

\textbf{Social Sciences \& Humanities}: educational leadership (2222), higher education (1560), management (1452), organizational behavior (989), clinical psychology (952), school administration (826), educational technology (806), education (757), teacher education (752), educational psychology (747), psychology (746), secondary education (730), womens studies (726), elementary education (600), health care management (585), education policy (571), educational evaluation (562), curriculum development (546), social psychology (538), special education (523), business administration (521), behavioral psychology (487), higher education administration (473), educational administration (471), african american studies (471), counseling psychology (451), public policy (449), adult education (449), music (415), community college education (398), health education (384), occupational psychology (374), individual \& family studies (374), organization theory (374), gender studies (369), political science (357), cognitive psychology (355), educational tests \& measurements (351), economics (345), communication (341), mathematics education (341), developmental psychology (337), social research (332), public administration (321), reading instruction (321), cultural anthropology (320), educational sociology (314), middle school education (308), social work (303), linguistics (295), science education (291), religion (286), early childhood education (282), ethnic studies (279), black studies (276), philosophy (267), spirituality (260), criminology (259), multicultural education (256), pedagogy (247), english as a second language (247), hispanic american studies (233), military studies (232), sociology (225), american history (225), literacy (219), instructional design (207), lgbtq studies (199), music education (196), finance (191), school counseling (170), religious education (170), language arts (170), personality psychology (168), marketing (165), business education (165), theology (163), native american studies (163), behavioral sciences (151), asian studies (148), web studies (148), international relations (143), archaeology (141), mass communications (141), social studies education (140), gerontology (138), entrepreneurship (133), bilingual education (132), latin american studies (131), american studies (128), ethics (126), education philosophy (121), american literature (121), education finance (119), law (117), british and irish literature (116), social structure (115), african studies (115), rhetoric (111), art history (111), accounting (109), labor relations (108), vocational education (107), quantitative psychology (99), religious history (95), language (93), history (92), art education (91), disability studies (89), economic theory (87), south asian studies (87), middle eastern studies (86), european history (86), labor economics (83), education history (82), asian american studies (77), film studies (76), comparative literature (75), speech therapy (74), multimedia communications (72), recreation (69), continuing education (65), business and secretarial schools (65), agricultural economics (64), latin american history (64), design (63), experimental psychology (62), biblical studies (62), demography (57), theater (57), sociolinguistics (56), performing arts (56), public health education (56), caribbean studies (55), science history (55), sports management (54), modern literature (54), biographies (54), banking (53), foreign language education (53), international law (53), physical education (53), literature (52), sub saharan africa studies (51), gifted education (51), commerce-business (50), law enforcement (50), judaic studies (49), foreign language (48), therapy (48), physical anthropology (47), classical studies (45), peace studies (45), modern history (44), philosophy of science (43), environmental education (41), clerical studies (40), journalism (39), museum studies (39), folklore (39), environmental economics (39), cultural resources management (37), fine arts (37), epistemology (37), latin american literature (36), romance literature (36), military history (35), aesthetics (34), ancient history (34), black history (33), technical communication (33), medical ethics (33), south african studies (32), pastoral counseling (32), psychotherapy (32), clergy (31), asian literature (30), creative writing (29), islamic studies (28), economic history (27), medieval literature (25), european studies (25), management consulting services (24), dance (24), middle eastern history (24), curricula (23), teaching (23), music history (23), occupational therapy (22), philosophy of religion (22), occupational safety (22), environmental justice (21), african history (21), musical composition (21), forensic anthropology (20), divinity (20), asian history (20), metaphysics (20), agricultural education (19), area planning and development (19), modern language (19), medieval history (18), music theory (18), comparative religion (17), middle eastern literature (17), cognitive therapy (17), near eastern studies (16), russian history (15), holocaust studies (15), theater history (15), morphology (15), range management (14), east european studies (14), alternative dispute resolution (14), regional studies (14), slavic literature (13), industrial arts education (13), art criticism (13), pacific rim studies (13), african literature (13), slavic studies (13), behaviorial sciences (13), performing arts education (12), home economics education (12), environmental philosophy (11), public finance activities (11), germanic literature (11), families \& family life (11), personal relationships (11), academic guidance counseling (11), international affairs (10), world history (10), caribbean literature (10), music therapy (10), north african studies (9), ancient languages (9), canadian studies (9), german literature (8), administration of education programs (8), arts management (8), armed forces (8), experiments (8), mass media (8), composition (8), translation studies (8), southeast asian studies (8), fashion (7), area planning \& development (7), experimental/theoretical (6), administration of general economic programs (6), community colleges (6), intellectual property (5), junior colleges (5), icelandic \& scandinavian literature (5), engineering services (5), canadian history (5), restaurants and other eating places (4), managerial skills (4), canadian literature (4), organizational structure (4), home economics (4), monetary authorities-central bank (4), electronic shopping and mail-order houses (4), personality (4), french literature (4), musical performances (3), cinematography (3), united states (3), religious organizations (3), commercial banking (3), social trends \& culture (3), capital \& debt management (3), business associations (3), tax preparation (3), bookkeeping (3), and payroll services (3), scandinavian studies (3), native americans (3), native studies (3), minority \& ethnic groups (3), canon law (2), social policy (2), schools and educational services (2), classical literature (2), document preparation services (2), credit bureaus (2), environmental consulting services (2), small business (2), administration of urban planning and community and rural development (2), regulation (2), licensing (2), and inspection of miscellaneous commercial sectors (2), business to business electronic markets (2), direct life (2), and medical insurance carriers (2), colleges (2), universities (2), and professional schools (2), supermarkets and other grocery (except convenience) stores (2), interior design (2), middle ages (2), motion pictures (2), company specific (1), fine arts schools (1), patent law (1), acquisitions \& mergers (1), service industries not elsewhere classified (1), telephone call centers (1), public relations (1), transportation equipment industry (1), french canadian literature (1), french canadian culture (1), national security (1), history of oceania (1), africa (1), multinational corporations (1), media buying agencies (1), counseling education (1), multilingual education (1), direct mail advertising (1), other direct selling establishments (1), psychological tests (1), demographics (1), asia \& the pacific (1), boards of directors (1), hispanic americans (1), hotels (except casino hotels) and motels (1), all other schools and instruction (1), museums (1), civic and social organizations (1), educational support services (1), industrial design services (1), labor unions and similar labor organizations (1), theater companies and dinner theaters (1), residential mental health and substance abuse facilities (1), italian literature (1), environmental law (1)
}

\newpage
\section{Regression Analysis} \label{secSI:regression_analysis}
\begin{figure}[htp]
    \centering
    \includegraphics[width=16cm]{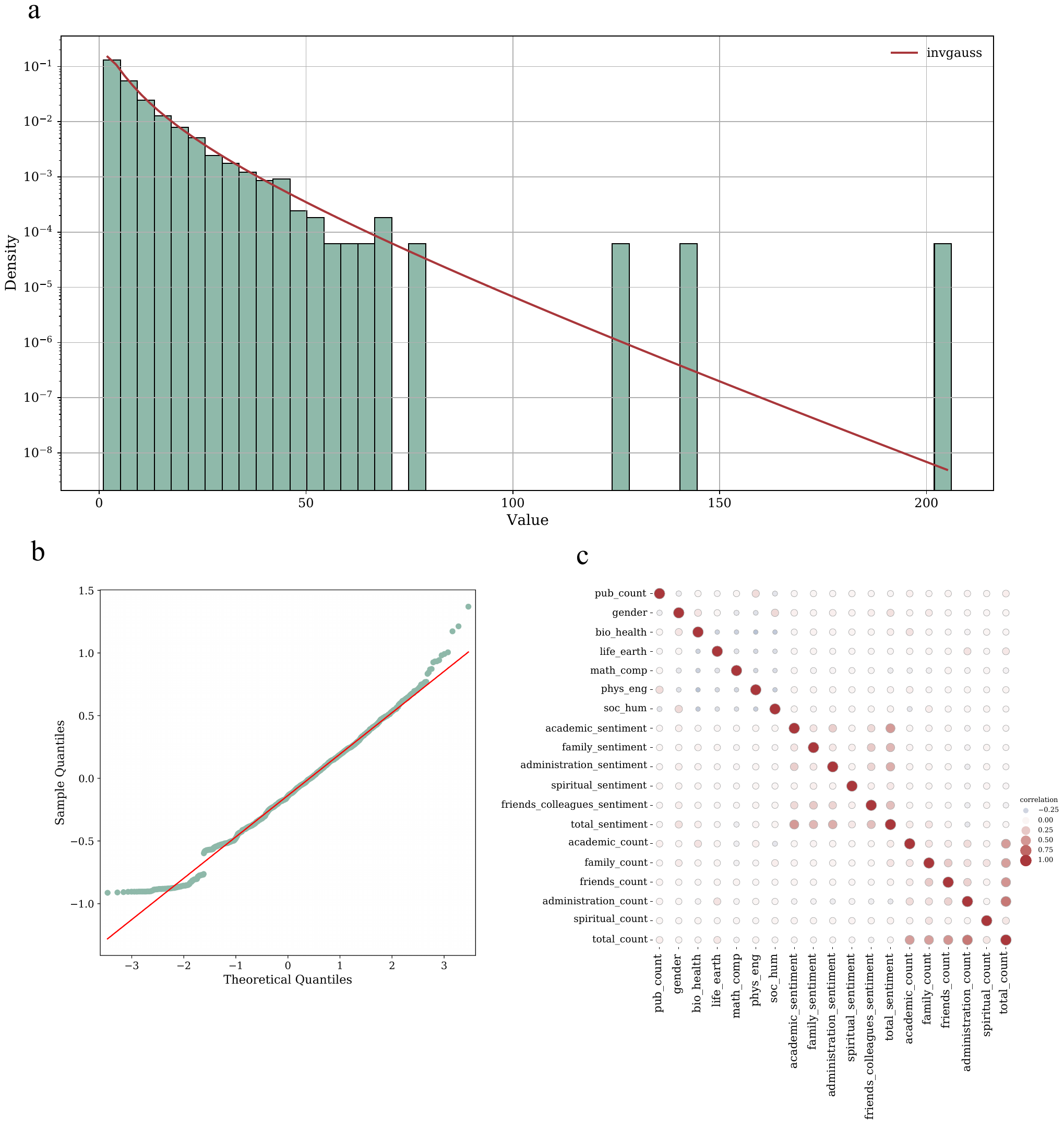}
    \caption{\textbf{Distribution Identification and Independent Variable Selection} Ten different distributions were fitted on the Publication Count histogram and the best distribution was plotted (a). Before checking the variation inflation factor (VIF), the correlations between independent variables were visually represented (b). While the points fall along a line in the middle of the graph, they curve off in the extremities (c).}
    \label{fig:regression-prep}
\end{figure}

A regression analysis is used to estimate the effect of independent variables on the target variable.
We used generalized linear model (GLM) with Inverse Gaussian distribution since linearity and normality assumptions do not hold in our case. 
The histogram that shows the publication counts and Inverse Gaussian distribution can be seen on Fig.\ref{fig:regression-prep}(a). A QQ-plot is also given on Fig.\ref{fig:regression-prep}(b) to motivate our preference using this model. 

After selecting the regression model, to detect multicollinearity and select the variables that are going to be used in regression analysis, we checked the variation inflation factor (VIF) and removed those which had higher than 10. The remaining covariates were included as given in the Fig.\ref{fig:regression-prep}.

\newpage
\section{University Rankings} 
\label{secSI:uni_rankings}

\begin{figure}[htp]
    \centering
    \includegraphics[width=17.5cm]{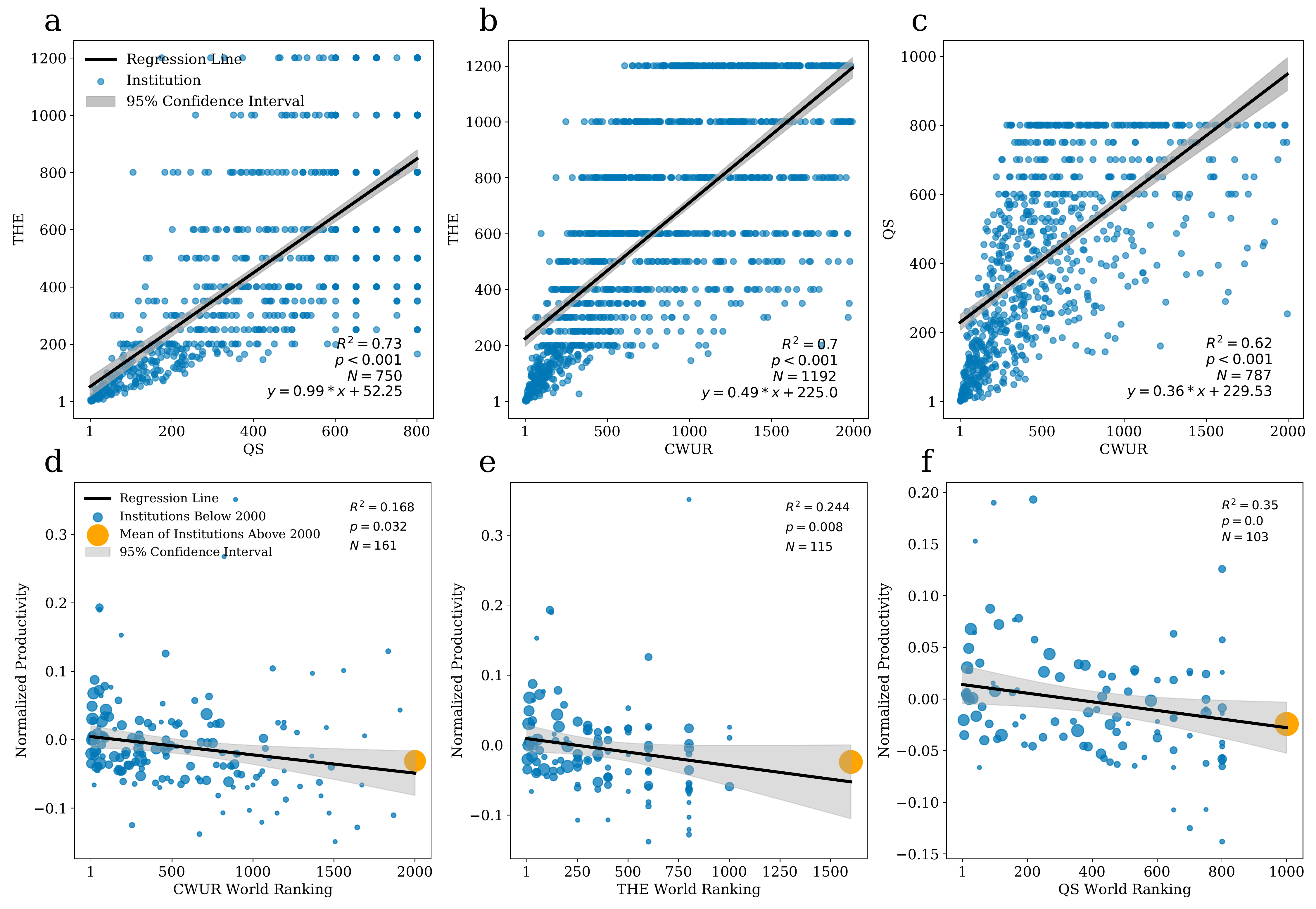}
    \caption{\textbf{Comparison of Different University Ranking Lists and Correlation with Productivity Levels} Relationship between QS and THE rankings (a), CWUR and THE rankings (b), CWUR and QS rankings (c). Relationship between normalized productivity and CWUR (d), THE (e), QS (f).
    N represents the number of institutions that match for both lists while the equations are associated with the given regression lines in figures. R squared values denote Spearman's rank correlation.}
    \label{fig:qs_the_cwur}
\end{figure}

We analyzed rankings of university affiliations of doctoral students with their performance and academic support networks. However, since nomenclatures for same universities may change from one database to another, we had to match ranking lists and university names in our PhD acknowledgement database manually before conducting a research on this part.
We investigated three widely used world ranking lists for universities: Center for World University Rankings (CWUR), Times Higher Education (THE), and Quacquarelli Symonds (QS), which included 2000, 1663 and 1000 universities, respectively. As can be seen on Fig.\ref{fig:qs_the_cwur}(a-c), pairwise correlation between these three lists can be considered as at least moderate with a minimum value of 62\% for CWUR-QS pair. Therefore, we have concluded that they are consistent enough to make an analysis on. Hence, we checked whether they are correlated with the number of publications or not. Fig.\ref{fig:qs_the_cwur}(d-f) shows that universities with higher rankings have on average PhD students with higher productivity levels.

\begin{figure}[htp]
    \centering
    \includegraphics[width=16cm]{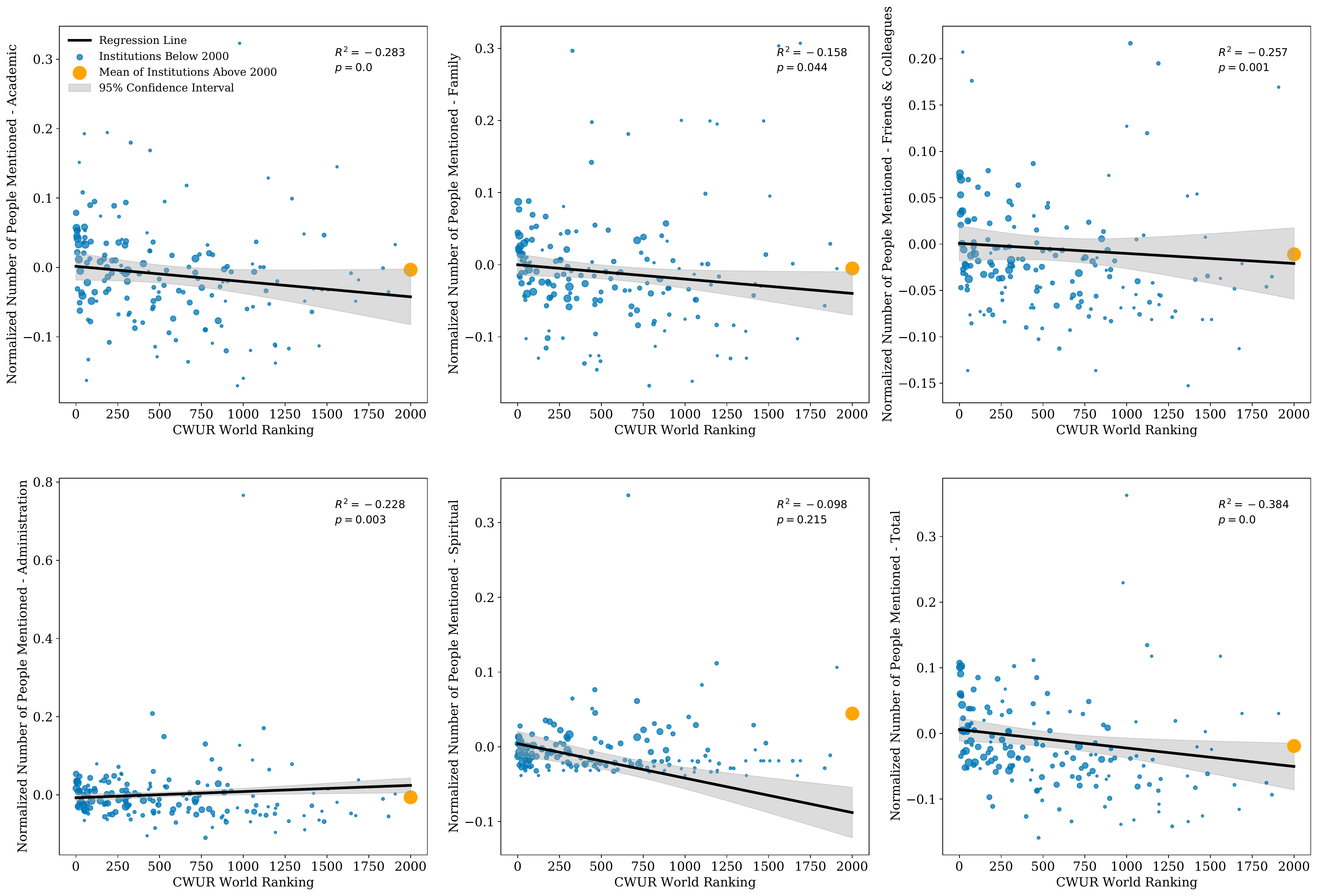}
    \caption{\textbf{Total Number of Mentions and CWUR World Rankings}}
    \label{fig:cwur_counts}
\end{figure}

\begin{figure}[htp]
    \centering
    \includegraphics[width=16cm]{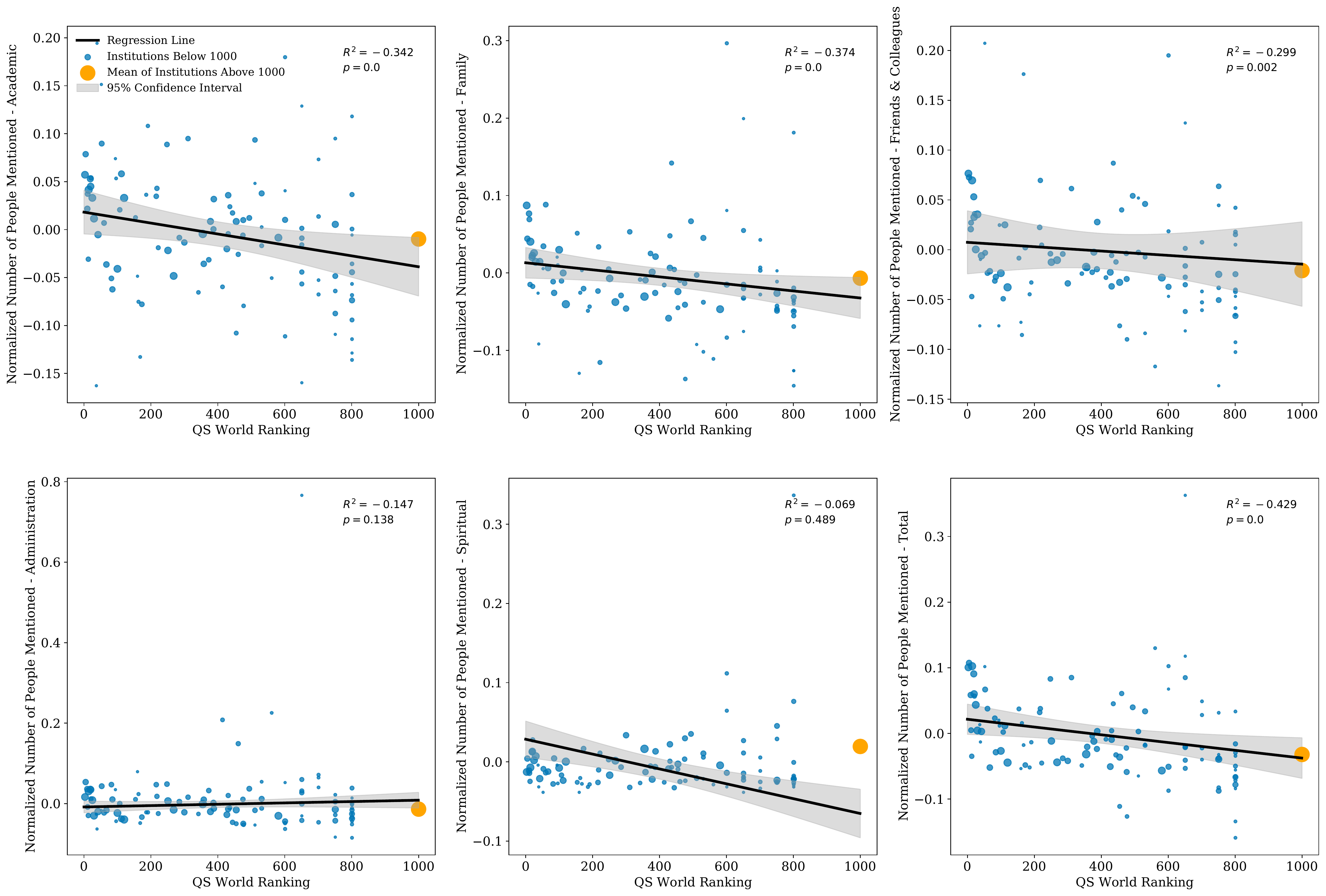}
    \caption{\textbf{Total Number of Mentions and QS World Rankings}}
    \label{fig:qs_counts}
\end{figure}

\begin{figure}[htp]
    \centering
    \includegraphics[width=16cm]{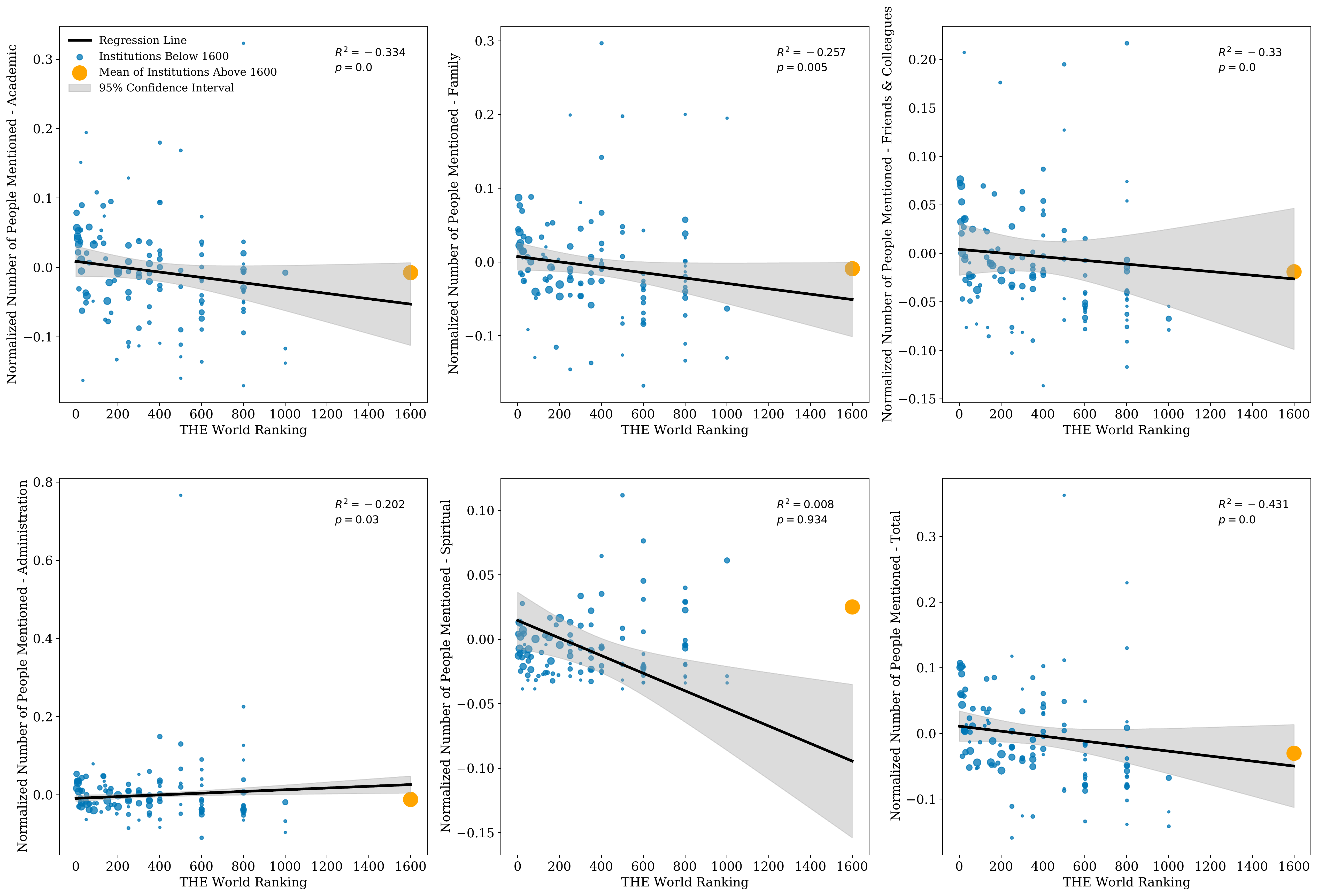}
    \caption{\textbf{Total Number of Mentions and THE World Rankings}}
    \label{fig:the_counts}
\end{figure}

Our results have shown that number of academic mentions and total mentions have a positive correlation with university rankings for each of the ranking lists as well. And, when we normalized the number of mentions regarding students' gender and discipline, the analyses yielded the same results as shown in Fig.\ref{fig:cwur_counts}, Fig.\ref{fig:qs_counts} and Fig.\ref{fig:the_counts}.

\begin{figure}[htp]
    \centering
    \includegraphics[width=16cm]{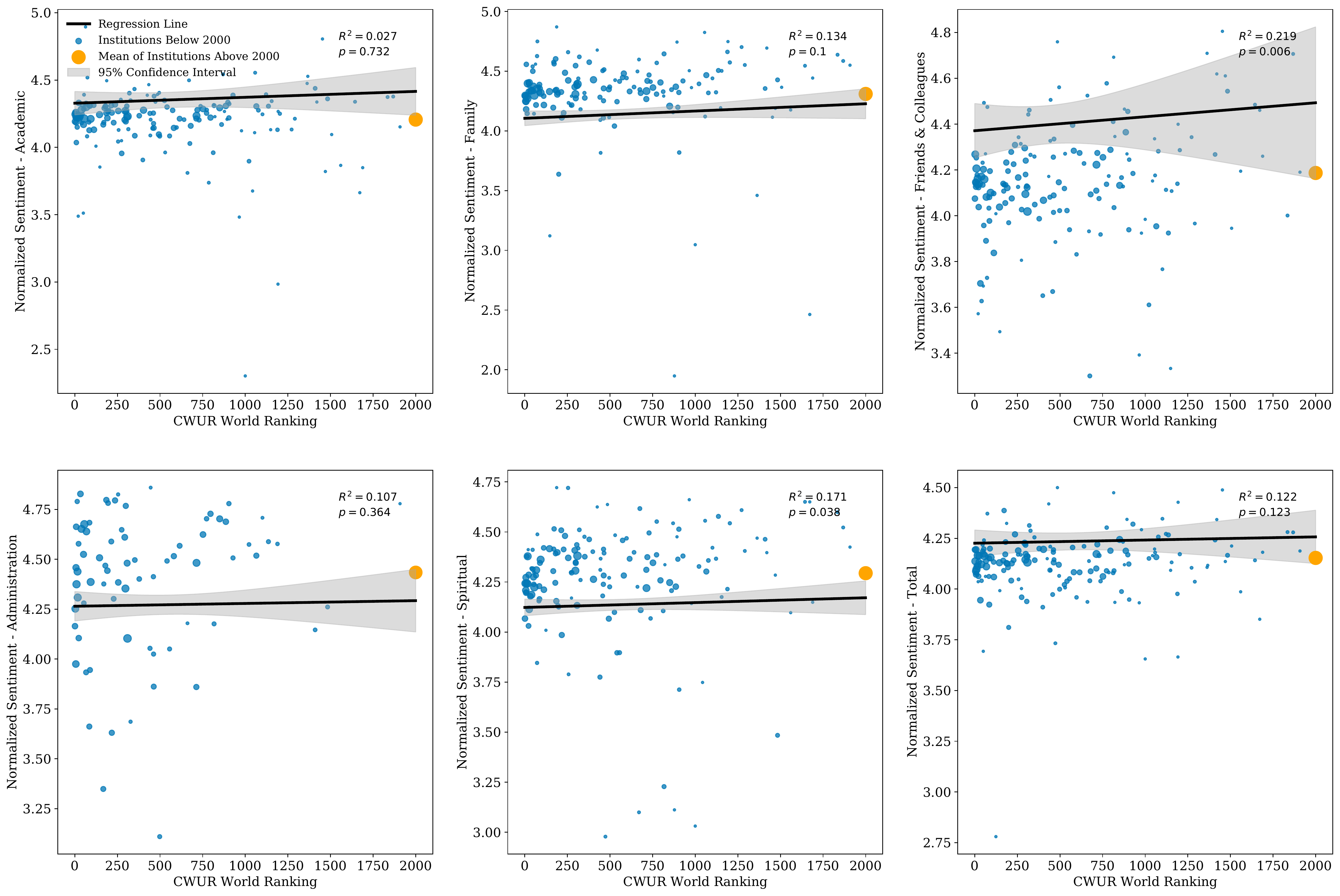}
    \caption{\textbf{Sentiment Scores for Each Support Provider Category and CWUR World Rankings}}
    \label{fig:cwur_sentiments}
\end{figure}

\begin{figure}[htp]
    \centering
    \includegraphics[width=16cm]{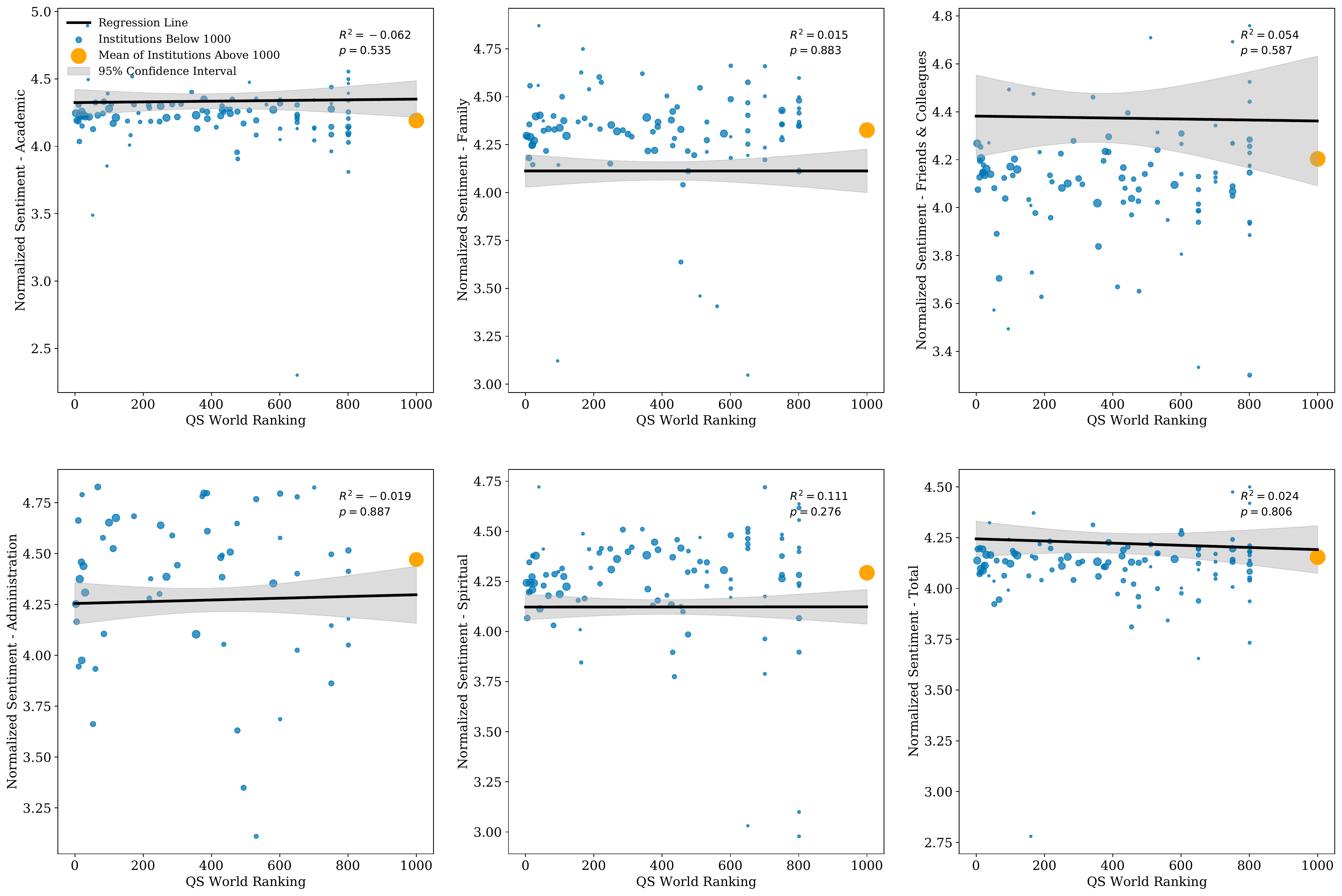}
    \caption{\textbf{Sentiment Scores for Each Support Provider Category and QS World Rankings}}
    \label{fig:qs_sentiments}
\end{figure}

\begin{figure}[htp]
    \centering
    \includegraphics[width=16cm]{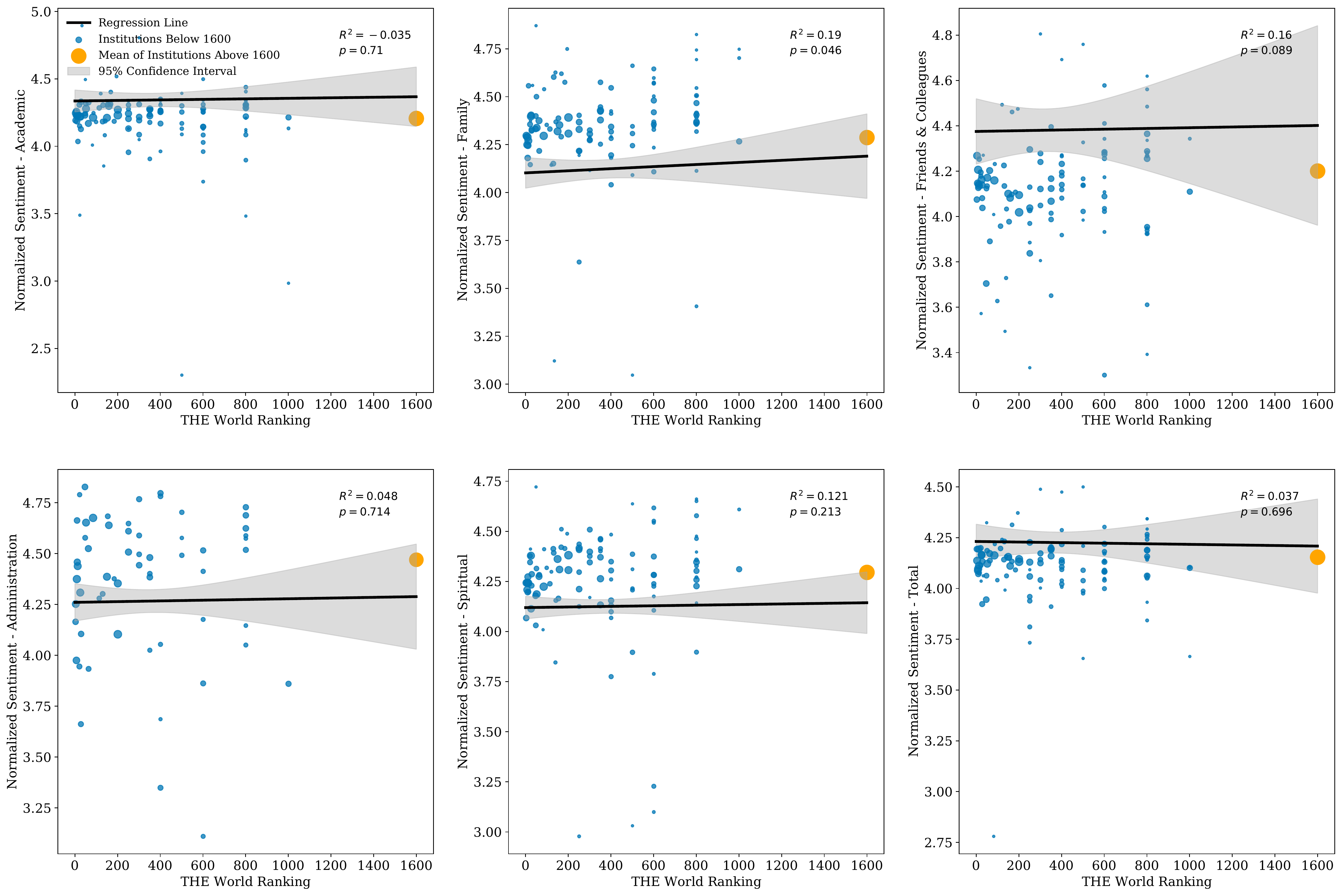}
    \caption{\textbf{Sentiment Scores for Each Support Provider Category and THE World Rankings}}
    \label{fig:the_sentiments}
\end{figure}

We examined whether there is a significant relationship between university rankings and sentiment scores as well. The results have shown that none of the sentiment scores can be explained by university rankings (see Fig.\ref{fig:cwur_sentiments}, \ref{fig:qs_sentiments}, \ref{fig:the_sentiments}).

\end{document}